\documentclass[journal]{IEEEtran}

\usepackage{booktabs}

\usepackage{amsmath,epsfig,amssymb,amsthm,url,amsfonts}
\usepackage{algorithm}
\usepackage{algpseudocode}
\usepackage{multirow}
\usepackage{mdframed}
\usepackage{url}
\usepackage{enumitem}
\usepackage[makeroom]{cancel}
\usepackage{dblfloatfix}
\usepackage{array}
\usepackage{epstopdf}
\usepackage{float}
\usepackage{subfig}
\usepackage{breqn}
\usepackage{cite}
\usepackage{xcolor}
\usepackage{tabularx}
\usepackage[printonlyused,withpage]{acronym}
\usepackage{balance}
\usepackage{float}
\setlist[itemize]{label=$\triangleright$}
\newtheoremstyle{break}
{}
{}
{\itshape}
{}
{\bfseries}
{.}
{\newline}
{}
\theoremstyle{break}

\theoremstyle{definition}

\DeclareMathOperator{\vect}{Vect}

\newcommand{\E}{\mathbb{E}}
\newcommand{\bs}{\boldsymbol}

\def\mbf#1{\mathbf{#1}}

\def\mbb#1{\mathbb{#1}}
\def\bs#1{\boldsymbol{#1}}

\makeatletter
\def\thmhead@plain#1#2#3{%
	\thmname{#1}\thmnumber{\@ifnotempty{#1}{ }\@upn{#2}}%
	\thmnote{ {\the\thm@notefont#3}}}
\let\thmhead\thmhead@plain
\makeatother

\newcommand{\argmax}{\operatornamewithlimits{argmax}}

\newcommand{\lk}{ \left\{ }
\newcommand{\rk}{ \right\} }

\newcommand{\Rbb}{{\mathbb{R}}}

\newcommand{\Hb}{{\bf H}}

\newcommand{\sbb}{{\bf s}}

\newcommand{\zb}{{\bf z}}

\newcommand{\Wb}{{\bf W}}

\newcommand{\cc}{{\cal c}}

\newcommand{\gammab}{{\mbox{\boldmath $\gamma$}}}

\newcommand{\vb}{{\mathbf v}}

\newsavebox\mybox

\acrodef{SE}{speech enhancement}
\acrodef{STOI}{short-time objective intelligibility}
\acrodef{STFT}{short-time Fourier transform}
\acrodef{PSD}{power spectral density}
\acrodef{NMF}{nonnegative matrix factorization}
\acrodef{AV}{audio-visual}
\acrodef{DNN}{deep neural network}
\acrodefplural{DNNs}{deep neural networks}
\acrodef{VAE}{variational auto-encoder}
\acrodefplural{VAEs}{variational auto-encoders}
\acrodef{CVAE}{conditional variational auto-encoder}
\acrodefplural{CVAEs}{conditional variational auto-encoders}
\acrodef{A-VAE}{audio VAE}
\acrodef{V-VAE}{visual VAE}
\acrodef{AV-CVAE}{audio-visual CVAE}
\acrodef{ROI}{region of interest}
\acrodef{MCMC}{Markov Chain Monte Carlo}
\acrodef{EM}{expectation-maximization}
\acrodef{MCEM}{Monte Carlo expectation-maximization}
\acrodef{TF}{time frequency}
\acrodef{ELBO}{evidence lower bound}
\acrodef{ROI}{region of interest}
\acrodef{LR}{Living Room}
\acrodef{SDR}{signal-to-distortion ratio}
\acrodef{PESQ}{perceptual evaluation of speech quality}
\acrodef{ASE}{audio speech enhancement}
\acrodef{VSE}{visual speech enhancement}
\acrodef{AVSE}{audio-visual speech enhancement}
\acrodef{SNR}{signal-to-noise ratio}
\acrodefplural{SNRs}{signal-to-noise ratios}
\acrodef{LSTM}{long short-term memory}

\usepackage[final]{review}
\setrevision{2}

\begin{document}
\title{Audio-visual Speech Enhancement Using Conditional Variational Auto-Encoders}

\author{Mostafa Sadeghi,$^{1,2}$ Simon Leglaive,$^{1,3}$ Xavier Alameda-Pineda,$^{1,2}$ \textit{Senior Member, IEEE}, Laurent Girin,$^{1,2,4}$ and Radu Horaud$^{1,2}$
\thanks{$^1$Inria Grenoble Rh\^one-Alpes, Montbonnot Saint-Martin.}
\thanks{$^2$Univ. Grenoble Alpes, Grenoble, France.}
\thanks{$^3$CentraleSup\'{e}lec, IETR.}
\thanks{$^4$GIPSA-Lab, Saint-Martin d'H\`eres, France.}

\thanks{This work has been supported by the Multidisciplinary Institute in Artificial Intelligence (MIAI), Grenoble, France.}	
}


\maketitle

\begin{abstract}

Variational auto-encoders (VAEs) are deep generative latent variable models that can be used for learning the distribution of complex data. VAEs have been successfully used to learn a probabilistic prior over speech signals, which is then used to perform speech enhancement. One advantage of this generative approach is that it does not require pairs of clean and noisy speech signals at training. In this paper, we propose audio-visual variants of VAEs for single-channel and speaker-independent speech enhancement. We develop a conditional VAE (CVAE) where the audio speech generative process is conditioned on visual information of the lip region. At test time, the audio-visual speech generative model is combined with a noise model based on nonnegative matrix factorization, and speech enhancement relies on a Monte Carlo expectation-maximization algorithm. Experiments are conducted with the recently published NTCD-TIMIT dataset \addnote[grid-abs]{1}{as well as the GRID corpus}. The results confirm that the proposed audio-visual CVAE effectively fuses audio and visual information, and it improves the speech enhancement performance compared with the audio-only VAE model, especially when the speech signal is highly corrupted by noise. We also show that the proposed unsupervised audio-visual speech enhancement approach outperforms a state-of-the-art supervised deep learning method. 

\end{abstract}


\IEEEpeerreviewmaketitle

\section{Introduction}
\label{sec:introduction}

The problem of \ac{SE} consists in estimating clean-speech signals from noisy single-channel or multiple-channel audio recordings. There is a long tradition of \ac{ASE} methods and associated algorithms, software and systems, e.g. \cite{lim1983speech,benesty2006speech,loizou2007speech}. In this paper we address the problem of \ac{AVSE}: in addition to audio, we exploit the benefits of visual speech information available with video recordings of lip movements. The rationale of \ac{AVSE} is that, unlike audio information, visual information (lip movements) is not corrupted by acoustic perturbations, and hence visual information can help the speech enhancement process, in particular in the presence of audio signals with low \acp{SNR}.

Although it has been shown that the fusion of visual and audio information is beneficial for various speech perception tasks, e.g. \cite{sumby1954visual, erber1975auditory, macleod1987quantifying}, \ac{AVSE} has been far less investigated than \ac{ASE}. \ac{AVSE} methods can be traced back to \cite{girin1995noisy} and subsequent work, e.g. \cite{girin2001audio,fisher2001learning, deligne2002audio,goecke2002noisy,hershey2002audio,abdelaziz2013twin}. Not surprisingly, \ac{AVSE} has been recently addressed in the framework of \acp{DNN} and a number of interesting architectures and well-performing algorithms were developed, e.g. 
\cite{AfouCZ18, GabbSP18, GabbEHP18, hou2018audio, GogaAMBH18}.

In this paper we propose to fuse single-channel audio and single-camera visual information for speech enhancement in the framework of \acp{VAE}. This may well be viewed as a multimodal extension of \ac{VAE}-based methods of \cite{bando2018statistical,Leglaive_MLSP18,SekiguchiAPSIPA2018, Leglaive_ICASSP2019a,Leglaive_ICASSP2019b,PariDV19} which, up to our knowledge, yield state-of-the-art \ac{ASE} performance in an unsupervised learning setting. In order to incorporate visual observations into the \ac{VAE} speech enhancement framework, we propose to use 
\acp{CVAE} \cite{SohnLY15}. As in \cite{Leglaive_MLSP18} we proceed in three steps. 

First, the parameters of the \ac{AV-CVAE} architecture are learned using synchronized clean audio-speech and visual-speech data. This yields an audio-visual speech prior model. \addnote[unsup-int]{1}{The training is totally unsupervised, in the sense that speech signals mixed with various types of noise signal are not required. This stays in contrast with supervised \ac{DNN} methods that need to be trained in the presence of many noise types and noise levels in order to ensure generalization and good performance, e.g. \cite{AfouCZ18,GabbSP18,GabbEHP18,li2019multichannel}. } 
Second, the learned speech prior is used in conjunction with a mixture model and with a \ac{NMF} noise variance model, to infer both the gain, which models the time-varying loudness of the speech signal, and the \ac{NMF} parameters. Third, the clean speech is reconstructed using the speech prior (VAE parameters) as well as the inferred gain and noise variance. The latter may well be viewed as a probabilistic Wiener filter. 
The learned \ac{VAE} architecture and its variants, the gain- and noise- parameter inference algorithms, and the proposed speech reconstruction method are thoroughly tested and compared with a state-of-the-art method, using the NTCD-TIMIT dataset \cite{Abde17} \addnote[grid-int]{1}{as well as the GRID corpus} \cite{CookBCS06_grid} containing audio-visual recordings. 

The remainder of the paper is organized as follows. Section~\ref{sec:related} summarizes related work. In Section~\ref{sec:a_vae} we briefly review how to use a VAE to model the speech prior distribution. Then, in Section~\ref{sec:visual-vae} we introduce two VAE network variants for learning the speech prior from visual data. In Section~\ref{sec:av_vae}, we present the proposed AV-CVAE used to model the acoustic speech distribution conditioned by visual information. In Section~\ref{sec:inference}, we discuss the inference phase, i.e., the actual speech enhancement process. Finally, our experimental results are presented in Section~\ref{sec:exp}.\footnote{Supplementary materials with audio-visual and visual speech enhancement examples are provided at \url{https://team.inria.fr/perception/research/av-vae-se/}.}


\section{Related Work}
\label{sec:related}

Speech enhancement has been an extremely investigated topic for the last decades and a complete state of the art is beyond the scope of this paper. We briefly review the literature on single-channel \ac{SE} and then we discuss the most significant work in \ac{AVSE}. 

Classical methods use spectral subtraction \cite{boll1979suppression} and Wiener filtering \cite{lim1979enhancement} based on noise and/or speech \ac{PSD} estimation in the \ac{STFT} domain. Another popular family of methods is the short-term spectral amplitude estimator \cite{ephraim1984speech}, initially based on a local complex-valued Gaussian model of the speech \ac{STFT} coefficients  and then extended to other density models \cite{martin2005speech, erkelens2007minimum}, and to a log-spectral amplitude estimator \cite{ephraim1985speech, cohen2001speech}. A popular technique for modeling the \ac{PSD} of speech signals \cite{ISNMF} is \ac{NMF}, e.g., \cite{wilson2008speech, raj2011phoneme, mohammadiha2013supervised}.

More recently, \ac{SE} has been addressed in the framework of \acp{DNN} \cite{wang2018supervised}. \addnote[unsup-int2]{1}{Supervised methods learn mappings between noisy-speech and clean-speech spectrograms, which are then used to reconstruct a speech waveform
\cite{lu2013speech,xu2015regression,fu2016snr}. Alternatively, the noisy input is mapped onto a \ac{TF} mask, which is then applied to the input to remove noise and to preserve speech 
information as much as possible \cite{wang2013towards,wang2014training,weninger2015speech,li2019multichannel}. 
In order for these supervised learning methods to generalize well and to yield state-of-the-art results, the training data must contain 
a large variability in terms of speakers and, even more critically, in terms of noise types and noise levels \cite{wang2013towards,xu2015regression}; in practice this leads to cumbersome learning processes. } 
 \addnote[unsup]{2}{Alternatively, generative (or unsupervised) DNNs do not use any kind of noise information for training, and for this reason they are very interesting because they have very good generalization capabilities}. An interesting generative formulation is provided by \acp{VAE} \cite{KingRMW14}. Combined with \ac{NMF}, \ac{VAE}-based methods yield state-of-the-art \ac{SE} performance \cite{bando2018statistical,Leglaive_MLSP18,SekiguchiAPSIPA2018, Leglaive_ICASSP2019a,Leglaive_ICASSP2019b,PariDV19} for an unsupervised learning setting. VAEs conditioned on the speaker identity have also been used for speaker-dependent multi-microphone speech separation \cite{KameLIM19, LiICASSP2019} and dereverberation \cite{InoueICASSP2019}.

The use of visual cues to complement audio, whenever the latter is noisy, ambiguous or incomplete, has been thoroughly studied in psychophysics \cite{sumby1954visual, erber1975auditory, macleod1987quantifying}. Indeed, speech production implies simultaneous air circulation through the vocal tract and tongue and lip movements, and hence speech perception is multimodal. Several computational models were proposed to exploit the correlation between audio and visual information for the perception of speech, e.g. \cite{fisher2001learning,hershey2002audio}.
A multi-layer perceptron architecture was proposed in  \cite{girin2001audio} to map noisy-speech linear prediction features concatenated with visual features onto clean-speech linear prediction features. Then Wiener filters were built for denoising. 
Audio-visual Wiener filtering was later extended using phoneme-specific Gaussian mixture regression and filterbank 
audio features \cite{almajai2010visually} . Other \ac{AVSE} methods exploit noise-free visual information   \cite{deligne2002audio,
goecke2002noisy} or make use of twin hidden Markov models (HMMs) \cite{abdelaziz2013twin}. 

State-of-the-art supervised \ac{AVSE} methods are based on 
\acp{DNN}. The rationale of \cite{AfouCZ18, GabbEHP18} is to use visual information to predict a \ac{TF} \addnote[text-1]{1}{soft }mask in the \ac{STFT} domain 
and to apply this mask to the audio input in order to remove noise. In \cite{GabbEHP18} a video-to-speech architecture is trained for each speaker in the dataset, which yields a speaker-dependent \ac{AVSE} method. The architecture of \cite{AfouCZ18} is composed of a magnitude subnetwork that takes both visual and audio data as inputs, and a phase subnetwork that only takes audio as input. Both subnetworks are trained using ground-truth clean speech. Then, the magnitude subnetwork predicts a binary mask which is then applied to both the magnitude and phase spectrograms of the input signal, thus predicting a filtered speech spectrogram. The architectures of \cite{hou2018audio} and \cite{GabbSP18} are quite similar: they are composed of two subnetworks, one for processing noisy speech and one for processing visual speech. The two encodings are then concatenated and processed to eventually obtain an enhanced speech spectrogram. The main difference between \cite{hou2018audio} and \cite{GabbSP18} is that the former predicts both enhanced visual and audio speech, while the latter predicts only audio speech. The idea of obtaining a binary mask for separating speech of an unseen speaker from an unknown noise was exploited in 
\cite{GogaAMBH18}: a hybrid \ac{DNN} model integrates a stacked \ac{LSTM} and convolutional \ac{LSTM} for \ac{AV} mask estimation.

In the supervised deep learning methods just mentioned, generalization to unseen data is a critical issue. The major issues are noise and speaker variability. Therefore, training these methods requires noisy mixtures with a large number of noise types and speakers, in order to guarantee generalization. In comparison, the proposed method is totally unsupervised: its training is based on \acp{VAE} and it only requires clean audio speech and visual speech. The gain and the noise variance are estimated at testing using a \ac{MCEM} algorithm \cite{wei1990monte}. The clean speech is then reconstructed from the audio and visual inputs using the learned parameters. The latter may well be viewed as a probabilistic Wiener filter. This stays in contrast with the vast majority of supervised \ac{DNN}-based \ac{AVSE} methods that predict a \ac{TF} mask which is applied to the noisy input. Empirical validation, based on standard \ac{SE} scores and using a widely used publicly available dataset, shows that our method outperforms the \ac{ASE} method \cite{Leglaive_MLSP18} as well as the state-of-the-art supervised \ac{AVSE} method \cite{GabbSP18}.

\section{Audio VAE}
\label{sec:a_vae}

In this section, we briefly review the deep generative speech model that was first proposed in \cite{bando2018statistical} along with its parameters estimation procedure using VAEs \cite{KingRMW14}. Let $s_{fn}$ denote the complex-valued speech \ac{STFT} coefficient at frequency index $f \in \{0,...,F-1\}$ and at frame index $n$. At each \ac{TF} bin, we have the following model which will be referred to as \ac{A-VAE}:
%
%
\begin{align}
\label{decoder_VAE}
s_{fn} | \mathbf{z}_n &\sim \mathcal{N}_c(0, \sigma_{f}(\mathbf{z}_n)), \\
\label{prior_VAE}
\mathbf{z}_n &\sim \mathcal{N}(\mathbf{0}, \mathbf{I}) 
\end{align}
where $\mathbf{z}_n \in \mathbb{R}^L$, with $L \ll F$, is a latent random variable describing a speech generative process, 
$\mathcal{N}(\mathbf{0}, \mathbf{I})$ is a zero-mean multivariate Gaussian distribution with identity covariance matrix, and 
 $\mathcal{N}_c(0, \sigma)$ is a
univariate complex proper Gaussian distribution with zero mean and variance $\sigma$. Let $\mathbf{s}_n\in \mathbb{C}^F$ be the vector whose components are the speech \ac{STFT} coefficients at frame $n$. The set of non-linear functions $\{\sigma_{f}: \mbb{R}^L \mapsto 
\mbb{R}_+\}_{f=0}^{F-1}$ are modeled as neural networks  sharing the input $\mbf{z}_n \in 
\mathbb{R}^L$. The parameters of these neural networks are collectively denoted with $\bs{\theta}$. This variance can be interpreted as a model for the short-term \ac{PSD} of the speech signal.

\begin{figure*}[t!h!]
	\centering
	{\includegraphics[width=0.7\textwidth]{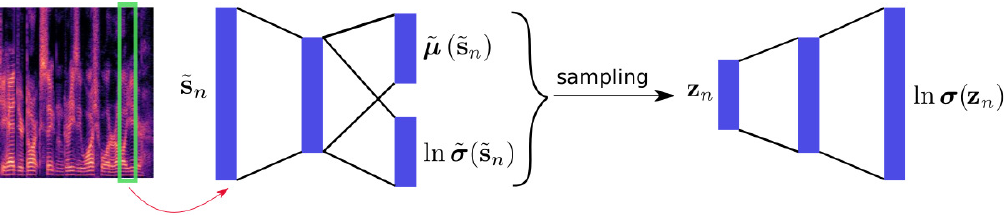}}
	\caption{\label{fig:vae} The A-VAE network used for learning a speech prior using audio data. The encoder network (left) takes as input the squared magnitude vector $\tilde{\sbb}_n $, associated with the STFT frame $\sbb_n$ (outlined in green), and outputs the mean and variance of the posterior distribution $ q\left(\mathbf{z}_n | {\sbb}_n ; \bs{\psi}\right) $. The decoder network (right) takes $ \zb_n $ as input (sampled from the posterior distribution) and outputs the variance of $ p(\sbb_n|\zb_n;\bs{\theta}) $. }
\end{figure*}

An important property of \acp{VAE} is to provide an efficient way of learning the parameters $\bs{\theta}$ of such generative models \cite{KingRMW14}, taking ideas from variational inference \cite{jordan1999introduction,blei2017variational}. Let 
$\mbf{s} = \{ \mbf{s}_n \in \mbb{C}^F \}_{n=0}^{N_{tr}-1}$ be a training dataset of clean-speech STFT frames and let $\mbf{z} = \{ \mbf{z}_n \in 
\mbb{R}^L \}_{n=0}^{N_{tr}-1}$ be the associated latent variables. In the \ac{VAE} framework, the parameters $\bs{\theta}$ are estimated
by maximizing a lower bound of the log-likelihood, $\ln p(\mbf{s} ; \bs{\theta})$, called \ac{ELBO}, defined 
by:
\begin{equation}
\mathcal{L}\left(\sbb; \bs{\theta}, \bs{\psi}\right) \hspace{-1.5pt}=\hspace{-1.5pt}  \E_{q\left(\mathbf{z} | \mathbf{s} ; 
\bs{\psi}\right)}\left[ \ln p\left(\mathbf{s} | \mathbf{z} ; \bs{\theta} \right) \right] - D_{\text{KL}}\left(q\left(\mathbf{z} | \mathbf{s} 
; \bs{\psi}\right) \parallel p(\mathbf{z})\right),
\label{variational_bound}
\end{equation}
where $q\left(\mathbf{z}|\mathbf{s} ; \bs{\psi}\right)$ denotes an approximation of the intractable true posterior distribution 
$p(\mathbf{z} | \mathbf{s} ; \bs{\theta} )$, $ p(\mathbf{z}) $ is the prior distribution of $ \zb $, and $D_{\text{KL}}(q \parallel p) = 
\mathbb{E}_q[\ln(q/p)]$ is the Kullback-Leibler divergence. Independently, for all $l \in \{0,...,L-1\}$ and all $n \in \{0,...,N_{tr}-1\}$, $q(\mathbf{z} | \mathbf{s} ; \bs{\psi})$ is defined by:
\begin{equation}
z_{ln} | \mathbf{s}_n \sim \mathcal{N}\left(\tilde{\mu}_l\left(\tilde{\mathbf{s}}_n\right), 
\tilde{\sigma}_l\left(\tilde{\mathbf{s}}_n\right) \right),
\label{encoder_VAE}
\end{equation}
where $ \tilde{\mathbf{s}}_n \triangleq (|s_{0n}|^2 \dots |s_{F-1\: n}|^2)^{\top}$. The non-linear functions
$\{\tilde{\mu}_l: \mathbb{R}_+^{F} \mapsto \mathbb{R}\}_{l=0}^{L-1}$ and $\{\tilde{\sigma}_l: \mathbb{R}_+^{F} \mapsto 
\mathbb{R}_+\}_{l=0}^{L-1}$ are modeled as neural networks, sharing as input the speech power spectrum frame $ \tilde{\mathbf{s}}_n$, and collectively parameterized by $\bs{\psi}$. The parameter set  $\bs{\psi}$ is also estimated by maximizing the 
\emph{variational lower bound} defined in \eqref{variational_bound}, which is actually equivalent to minimizing the Kullback-Leibler divergence between $q\left(\mathbf{z}|\mathbf{s} ; \bs{\psi}\right)$ and the intractable true posterior distribution 
$p(\mathbf{z} | \mathbf{s} ; \bs{\theta} )$ \cite{jordan1999introduction}. Using \eqref{decoder_VAE}, \eqref{prior_VAE} and \eqref{encoder_VAE} we 
can develop this objective function as follows:
\begin{align}
\mathcal{L}\left(\sbb;\bs{\theta}, \bs{\psi}\right) \overset{c}{=}&- \sum_{f=0}^{F-1}\sum_{n=0}^{N_{tr}-1} \mathbb{E}_{q\left(\mathbf{z}_n | 
\mathbf{s}_n ; \bs{\psi} \right)}\left[ d_{\text{IS}}\left(\left|s_{fn}\right|^2 ; \sigma_f(\mathbf{z}_n)\right) \right] \nonumber \\
& \hspace{-1.0cm} + \frac{1}{2} \sum_{l=0}^{L-1} \sum_{n=0}^{N_{tr}-1}\left[ \ln \tilde{\sigma}_l\left(\tilde{\mathbf{s}}_n\right) - 
\tilde{\mu}^2_l\left(\tilde{\mathbf{s}}_n\right) - \tilde{\sigma}_l\left(\tilde{\mathbf{s}}_n\right) \right],
\label{ELBO}
\end{align}
where $d_{\text{IS}}(x;y) = x/y - \ln(x/y) - 1$ is the Itakura-Saito divergence \cite{ISNMF}. Finally, using sampling techniques combined with the 
so-called ``reparametrization trick'' \cite{KingRMW14} to approximate the intractable expectation in \eqref{ELBO}, one obtains an objective 
function which is differentiable with respect to both $\bs{\theta}$ and $\bs{\psi}$ and can be optimized using gradient-ascent 
algorithms \cite{KingRMW14}. 
The encoder-decoder architecture of the A-VAE speech prior is summarized in Figure~\ref{fig:vae}.

\begin{figure*}[t]
\centering
{\includegraphics[width=0.85\textwidth]{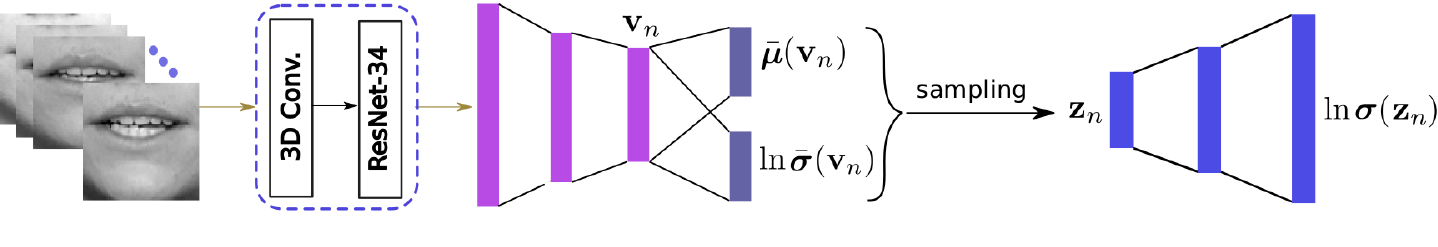}}
\caption{\label{fig:v_vae} The two V-VAE network variants (base and augmented) for learning speech prior from visual features. A lip \ac{ROI} is embedded into a visual feature vector, denoted $\mathbf{v}_n$, which is encoded and decoded using the same architecture and the same learning method as A-VAE. Optionally, one can also use a pre-trained network (dashed box) composed of a 3D convolution layer followed by a ResNet with 34 layers.}
\end{figure*}


\section{Visual VAE}
\label{sec:visual-vae}

We now introduce two VAE network variants for learning the speech prior from visual data, that will be referred to as \textit{base} \ac{V-VAE} and \textit{augmented} \ac{V-VAE}, and which are summarized in Figure~\ref{fig:v_vae}. As it can be seen, this architecture is similar to A-VAE, with the notable difference that it takes as input visual observations, namely lip images. In more detail, standard computer vision algorithms are used to extract a fixed-sized bounding-box from the image of a speaking face, with the lips in its center, i.e. a lip \ac{ROI}. This \ac{ROI} is embedded into a visual feature vector $\mathbf{v}_n\in\mathbb{R}^M$ using a two-layer fully-connected network, referred below as the \textit{base} network, where $M$ is the dimension of the visual embedding. Optionally, one can use an additional pre-trained \textit{front-end} network (dashed box) composed of a 3D convolution layer followed by a ResNet with 34 layers, as part of a network specifically trained for the task of supervised audio-visual speech recognition \cite{PetrSMC18}. This second option is referred to as \textit{augmented} V-VAE.

In variational inference \cite{jordan1999introduction,blei2017variational}, any distribution over the latent variables $\mathbf{z}$ can be considered for approximating the intractable posterior $p(\mathbf{z} | \mathbf{s} ; \bs{\theta} )$ and for defining the ELBO.
For the V-VAE model, we explore the use of an approximate posterior distribution $ q(\zb|\vb;\bs{\gamma}) $ defined by:
\begin{align}
\label{eq:latent_vvae}
z_{ln} | \vb_n \sim \mathcal{N}\left(\bar{\mu}_l(\vb_n), \bar{\sigma}_{l}(\vb_n)\right),
\end{align}
where $\mathbf{v}=\{\mathbf{v}_n\}_{n=1}^{N_{tr}-1}$ is the training set of visual features, and where
the non-linear functions $\{\bar{\mu}_l:\Rbb^M \mapsto \mathbb{R}\}_{l=0}^{L-1}$ and $\{\bar{\sigma}_l:\Rbb^M \mapsto \mathbb{R}_+\}_{l=0}^{L-1}$ are collectively modeled with a
neural network parameterized by $ \gammab $ and which takes $ \vb_n $ as input. Notice that V-VAE and A-VAE share the same decoder architecture, i.e. \eqref{decoder_VAE}. Eventually, the objective function of V-VAE has the same structure as \eqref{ELBO} and hence one can use the same gradient-ascent algorithm as above to estimate the parameters of the V-VAE network. 

\begin{figure*}[t]
	\centering
	{\includegraphics[width=0.7\textwidth]{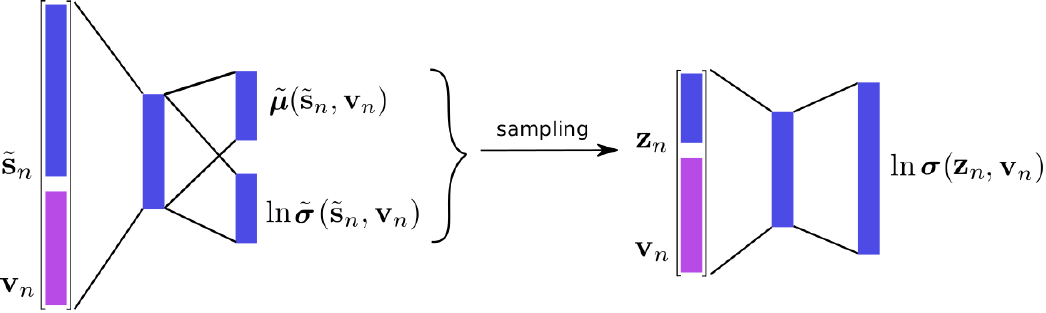}}
	\caption{\label{fig:vae_encdec} Pipeline of the proposed \ac{AV-CVAE} architecture for learning an audio-visual speech prior for  speech enhancement. The encoder takes a single frame of squared magnitude of speech's STFT, denoted by $ \tilde{\sbb}_n $, as well as the corresponding visual feature vector $ \vb_n $, and outputs the parameters of the posterior distribution $ q\left(\mathbf{z}_n | {\sbb}_n, \vb_n ; \bs{\psi}\right) $. The decoder network takes $ \zb_n $, sampled from the posterior distribution, together with $ \vb_n $ as input and outputs the variance of  $ p({\sbb}_n|\zb_n, \vb_n; \bs{\theta}) $.}
\end{figure*}

\section{Audio-visual VAE}
\label{sec:av_vae}
We now investigate an audio-visual \ac{VAE} model, namely a model that combines audio speech with visual speech. The rationale behind this multimodal approach is that audio data are often corrupted by noise while visual data are not.
Without loss of generality, it will be assumed that audio and visual data are synchronized, i.e. there is a video frame associated with each audio frame. 

In order to combine the above A-VAE and V-VAE formulations, we consider the \ac{CVAE} framework to learn structured-output representations \cite{SohnLY15}. At training, a \ac{CVAE} is provided with data as well as with associated class labels, such that the network is able to learn a structured data distribution. At test, the trained network is provided with a class label to generate samples from the corresponding class. \acp{CVAE} have been proven to be very effective for missing-value inference problems, e.g., computer vision problems with partially available input-output pairs \cite{SohnLY15}.


In the case of \ac{AV} speech enhancement we consider a training set of $N_{tr}$ synchronized frames of \ac{AV} features, namely $(\mathbf{s}, \mathbf{v})=\{\mathbf{s}_n, \mathbf{v}_n\}_{n=1}^{N_{tr}-1}$ where, as above, $\vb_n  \in \mathbb{R}^M$ is a lip \ac{ROI} embedding. The clean audio speech, which is only available at training, is conditioned on the observed visual speech. The visual information is however available both at training and at testing, therefore it serves as a deterministic prior on the desired clean audio speech. Interestingly, it also affects the prior 
distribution of $ \zb_n $. To summarize, the following latent space model is considered, independently for all $l \in \{0,...,L-1\}$ 
and all TF bins $(f,n)$:
\begin{align}
\label{eq:model_cvae}
s_{fn} | \mathbf{z}_n, \vb_n &\sim \mathcal{N}_c(0, \sigma_{f}(\mathbf{z}_n, \vb_n)),\\
\label{eq:latent_cvae}
z_{ln} | \vb_n &\sim \mathcal{N}\left(\bar{\mu}_l(\vb_n), \bar{\sigma}_{l}(\vb_n)\right),
\end{align}
where the non-linear functions $\{{\sigma}_f:\Rbb^L\times \Rbb^M \mapsto \mathbb{R}_+\}_{f=0}^{F-1}$ are modeled as a neural network parameterized by $\bs{\theta}$ and taking $ \zb_n $ and $ \vb_n $ as input, and where \eqref{eq:latent_cvae} is identical with \eqref{eq:latent_vvae} but the corresponding parameter set $\gammab$ will have different estimates, as explained below.
Also, notice that $\sigma_f$ in \eqref{decoder_VAE}
and in \eqref{eq:model_cvae} are different, but they both correspond to the \ac{PSD} of the generative speech model. This motivates the abuse of 
notation that holds through the paper.  The proposed architecture is referred to as \ac{AV-CVAE} and is shown in Fig.~\ref{fig:vae_encdec}. 
Compared to A-VAE of Section~\ref{sec:a_vae} and Figure~\ref{fig:vae} and with V-VAE of Section~\ref{sec:visual-vae} and Figure~\ref{fig:v_vae}, the mean and variance of the $ \zb_n $ prior distribution, are conditioned by visual inputs.

We introduce now the distribution $q\left(\mathbf{z}|\mathbf{s}, \vb ; \bs{\psi}\right)$, which approximates the intractable posterior distribution 
$p(\mathbf{z} | \mathbf{s}, \vb ; \bs{\theta} )$, defined, as above, independently for all $l \in \{0,...,L-1\}$ and all 
frames:
\begin{equation}
z_{ln} | \mathbf{s}_n, \vb_n \sim \mathcal{N}\left(\tilde{\mu}_l\left(\tilde{\mathbf{s}}_n, \vb_n\right), 
\tilde{\sigma}_l\left(\tilde{\mathbf{s}}_n, \vb_n\right) \right),
\label{encoder_CVAE}
\end{equation}
where the non-linear functions $\{\tilde{\mu}_l: \mathbb{R}_+^{F}\times \Rbb^M \mapsto \mathbb{R}\}_{l=0}^{L-1}$ and $\{\tilde{\sigma}_l: \mathbb{R}_+^{F}\times \Rbb^M \mapsto \mathbb{R}_+\}_{l=0}^{L-1}$ are collectively modeled as an encoder neural network, parameterized by $\bs{\psi}$, that takes as input the speech power spectrum and its associated visual feature vector, at each frame.  
The complete set of model parameters, i.e. $ \gammab $, $\bs{\theta}$ and $\bs{\psi}$, can be estimated by maximizing a lower 
bound of the conditional log-likelihood $\ln p(\mbf{s}| \vb ; \bs{\theta}, \gammab)$ over the training dataset, defined by:
\begin{align}
\label{eq:cvae_elbo}
\mathcal{L}_{\text{av-cvae}}\left(\sbb, \vb;\bs{\theta}, \bs{\psi}, \gammab\right) & =   
 \E_{q\left(\mathbf{z} | \mathbf{s}, 
\vb ; \bs{\psi}\right)}\left[ \ln p\left(\mathbf{s} | \mathbf{z}, \vb ; \bs{\theta} \right) \right] \\
& - D_{\text{KL}}\left(q\left(\mathbf{z} 
| \mathbf{s}, \vb ; \bs{\psi}\right) \parallel p(\mathbf{z}| \vb; \gammab)\right), \nonumber
\end{align}
where $\mbf{z} = \{ \mbf{z}_n \in \mbb{R}^L \}_{n=0}^{N_{tr}-1}$. 
This network architecture appears to be very effective for the task at hand. In fact, if 
one looks at the cost function in \eqref{eq:cvae_elbo}, it can be seen that the KL term achieves its optimal value for $ q\left(\mathbf{z} | 
\mathbf{s}, \vb ; \bs{\psi}\right) = p(\mathbf{z}| \vb;\gammab) $. By looking at the encoder of Fig.~\ref{fig:vae_encdec}, this can happen 
by ignoring the contribution of the audio input. Moreover, the first term in the cost function 
\eqref{eq:cvae_elbo} attempts to reconstruct as well as possible the audio speech vector at the output of the decoder. This can be done by 
using the audio vector in the input of the encoder as much as possible. This stays in contrast with the optimal behavior of the second 
term which tries to ignore the audio input. By minimizing the overall cost, the visual and audio information can be fused 
in the encoder. 

During the training of \ac{AV-CVAE}, the variable $ \zb_n $ is sampled from the approximate posterior modeled by the encoder, and it is then passed to the decoder. However, at testing only the decoder and prior networks are used while the encoder is discarded.
Hence, $ \zb_n $ is sampled from the prior network, which is basically different from the encoder network. The KL-divergence term in the cost function \eqref{eq:cvae_elbo} is responsible for reducing as much as possible the discrepancy between the recognition and prior networks. One can even control this by weighting the the KL-divergence term with
$ \beta>1 $:
\begin{align}
\label{eq:cvae_elbobeta}
\mathcal{L}_{\beta\text{-av-cvae}}\left(\sbb, \vb;\bs{\theta}, \bs{\psi}, \gammab\right) & =  
\E_{q\left(\mathbf{z} | \mathbf{s}, \vb ; 
\bs{\psi}\right)}\left[ \ln p\left(\mathbf{s} | \mathbf{z}, \vb ; \bs{\theta} \right) \right] \\
& - \beta D_{\text{KL}}\left(q\left(\mathbf{z} 
| \mathbf{s}, \vb ; \bs{\psi}\right) \parallel p(\mathbf{z}| \vb; \gammab)\right). \nonumber
\end{align}
This  was introduced in \cite{HiggMPBGBML17}, namely $ \beta $-VAE, and was shown to facilitate the
automated discovery of interpretable factorized latent representations. 
However, in the case of the proposed \ac{AV-CVAE} architecture, we follow a different stragety, proposed in \cite{SohnLY15}, in order to decrease the gap 
between the recognition and prior networks. As a consequence, the \ac{ELBO} defined in \eqref{eq:cvae_elbo} is modified as follows:
\begin{align}
\label{eq:cvae_elbotild}
\tilde{\mathcal{L}}_{\text{av-cvae}}\left(\sbb, \vb;\bs{\theta}, \bs{\psi}, \gammab\right) &= \alpha\mathcal{L}_{\text{av-cvae}}\left(\sbb, 
\vb;\bs{\theta}, \bs{\psi}, \gammab\right) \\
&+ (1-\alpha)\E_{p\left(\mathbf{z} | \vb ; \gammab\right)}\left[ \ln p\left(\mathbf{s} | 
\mathbf{z}, \vb ; \bs{\theta} \right) \right], \nonumber
\end{align}
where $ 0\le\alpha\le 1 $ is a trade-off parameter. Note that the original \ac{ELBO} is obtained by setting $ \alpha=1 $. The new term in the 
right-hand side of the above cost function is actually the original reconstruction cost in \eqref{eq:cvae_elbo} but with each $ \zb_n $ being 
sampled from the prior distribution, i.e.,  $ p\left(\mathbf{z}_n | \vb_n ; \gammab \right) $. In this way the prior network is forced to learn 
latent vectors that are suitable for reconstructing the corresponding speech frames. As it will be shown below, this 
method significantly improves the overall speech enhancement performance.

To develop the cost function in \eqref{eq:cvae_elbotild}, we note that the KL-divergence term admits a closed-form solution, because all the 
distributions are Gaussian. Furthermore, since the expectations with respect to the approximate posterior and prior of $ \zb_n $ are 
not tractable, we approximate them using Monte-Carlo estimations, as is usually done in practice. After some mathematical manipulations, one obtains the following cost function:
\begin{align}
\label{eq:cvae_elbotild2}
\tilde{\mathcal{L}}_{\text{av-cvae}}\big(\sbb, & \vb;\bs{\theta}, \bs{\psi}, 
\gammab \big)  \\
 = & \frac{1}{R}\sum_{r=1}^{R}\sum_{n=0}^{N_{tr}-1}\Big(\alpha\ln p(\mathbf{s}_n | \mathbf{z}_{n,1}^{(r)}, \vb_n ; \bs{\theta})  \nonumber \\
& + (1-\alpha)\ln p(\mathbf{s}_n | \mathbf{z}_{n,2}^{(r)}, \vb_n ; \bs{\theta})\Big) \nonumber \\
&+\frac{\alpha}{2}\sum_{l=0}^{L-1}\sum_{n=0}^{N_{tr}-1}\Big( \ln \frac{\tilde{\sigma}_l\left(\tilde{\mathbf{s}}_n, 
\vb_n\right)}{\bar{\sigma_{l}}(\vb_n)} \nonumber \\
& - \frac{\ln \tilde{\sigma}_l\left(\tilde{\mathbf{s}}_n, \vb_n\right)+\left(\tilde{\mu}_l\left(\tilde{\mathbf{s}}_n, \vb_n\right)-\bar{\mu_{l}}(\vb_n)\right)^2}{\bar{\sigma_{l}}(\vb_n)}\Big), \nonumber
\end{align}
where $ \mathbf{z}_{n,1}^{(r)}\sim q(\mathbf{z}_n | \mathbf{s}_n, \vb_n ; \bs{\psi}) $ and $ \mathbf{z}_{n,2}^{(r)}\sim p(\mathbf{z}_n | 
\vb_n ; \gammab) $. This cost function can be optimized in a similar way as with classical \acp{VAE}, namely by using the 
reparametrization trick together with a stochastic gradient-ascent algorithm. Notice that the reparameterization trick must be used twice, for $ \mathbf{z}_{n,1}^{(r)}$ and for $ \mathbf{z}_{n,2}^{(r)}$.

\section{AV-CVAE for  Speech Enhancement}
\label{sec:inference}

This section describes the speech enhancement algorithm based on the proposed AV-CVAE speech model. It is very similar to the algorithm that was proposed in \cite{Leglaive_MLSP18} for audio-only speech enhancement with VAE. The unsupervised noise model is first presented, followed by the mixture model, and by the proposed algorithm to estimate the parameters of the noise model. Finally, clean-speech inference procedure is described. Through this section, $ \vb = \lk \vb_n\rk_{n=0}^{N-1} $, $ \mathbf{s} = \lk \mathbf{s}_n\rk_{n=0}^{N-1} $ and $ 
\mathbf{z} = \lk \mathbf{z}_n\rk_{n=0}^{N-1} $ denote the test sets of visual features, clean-speech STFT features and latent vectors, respectively. These variables are associated with a noisy-speech test sequence of $N$ frames. One should notice that the test data are different than the training data used in the previous sections. The observed microphone (mixture) frames are denoted with $ \mathbf{x} = \lk 
\mathbf{x}_n\rk_{n=0}^{N-1} $.

\subsection{Unsupervised Noise Model}

As in \cite{Leglaive_MLSP18, bando2018statistical}, we use an unsupervised \ac{NMF}-based Gaussian noise model that 
assumes independence across \ac{TF} bins:
\begin{equation}
b_{fn} \sim \mathcal{N}_c\left(0, \left(\mathbf{W}_b\mathbf{H}_b\right)_{fn}\right),
\label{eq:noise-model}
\end{equation}
where $ \mathbf{W}_b\in\Rbb_+^{F\times K}$ is a nonnegative matrix of spectral power patterns and $\mathbf{H}_b\in\Rbb_+^{K\times N} $ is a nonnegative matrix of temporal activations, with $ K $ being chosen such that $ K(F + N ) \ll F N $ \cite{ISNMF}. We remind that $ \mathbf{W}_b $ and $\mathbf{H}_b$ need to be estimated from the observed microphone signal.

\subsection{Mixture Model} 

The observed mixture (microphone) signal is modeled as follows:
\begin{equation}
x_{fn} = \sqrt{g_n} s_{fn} + b_{fn},
\label{mixture}
\end{equation}
for all \ac{TF} bins $ (f,n) $, where $g_n \in \mathbb{R}_+$ represents a frame-dependent and frequency-independent gain, as suggested in 
\cite{Leglaive_MLSP18}. This gain provides robustness of the AV-CVAE model with respect to the possibly highly varying loudness of the 
speech signal across frames. Let us denote by $\mathbf{g} = (g_0 \dots g_{N-1})^\top$ the vector of gain parameters that 
must be estimated. The speech and noise signals are further assumed to be mutually independent, 
such that by combining \eqref{eq:model_cvae}, \eqref{eq:noise-model} and \eqref{mixture}, we obtain, for all \ac{TF} bins $(f,n)$:
\begin{equation}
x_{fn} | \mathbf{z}_n, \vb_n \sim \mathcal{N}_c\left(0, g_n \sigma_f(\mathbf{z}_n, \vb_n) + 
\left(\mathbf{W}_b\mathbf{H}_b\right)_{f,n}\right).
\label{likelihood}
\end{equation}
Let $\mathbf{x}_n\in \mathbb{C}^F$ be the vector whose components are the \ac{STFT} noisy mixture coefficients at frame $n$.

\subsection{Parameter Estimation}
\label{sec:inferencephase}
Having defined the speech generative model \eqref{eq:model_cvae} and the observed mixture model \eqref{likelihood}, the inference 
process requires to estimate the set of model parameters $\bs{\phi} = \{\mathbf{W}_b, \mathbf{H}_b, \mathbf{g}\}$ 
from the set of observed \ac{STFT} coefficients $\mathbf{x}$ and of observed visual features $ \vb $. Then, these parameters will be used to 
estimate the clean-speech \ac{STFT} coefficients. 
Since integration with respect to the latent variables is intractable, straightforward maximum likelihood estimation of $ \bs{\phi} $ 
is not possible. Alternatively, the latent-variable structure of the model can be exploited to derive an \ac{EM} 
algorithm \cite{dempster1977maximum}. Starting from an initial set of model parameters $\bs{\phi}^\star$, \ac{EM}
consists of iterating until convergence between:
\begin{itemize}[leftmargin=*]
	\item[-] E-step: Evaluate $Q(\bs{\phi}; \bs{\phi}^\star)\hspace{-0.06cm}=\hspace{-0.06cm} \mathbb{E}_{p(\mathbf{z} | \mathbf{x}, 
\vb ; \bs{\phi}^\star)} \hspace{-0.12cm} \left[ \ln p(\mathbf{x}, \mathbf{z}, \vb ; \bs{\phi}) \right]$;
	\item[-] M-step: Update $\bs{\phi}^\star \leftarrow \argmax_{\bs{\phi}} \, Q(\bs{\phi}; \bs{\phi}^\star)$.
\end{itemize}

\subsubsection{E-Step}
\label{subsec:EStep}

Because of the non-linear relation between the observations and the latent variables in \eqref{likelihood}, we cannot compute the posterior 
distribution ${p(\mathbf{z} | \mathbf{x}, \vb ; \bs{\phi}^\star)}$, and hence we cannot evaluate $Q(\bs{\phi}; \bs{\phi}^\star)$ 
analytically. As in \cite{Leglaive_MLSP18}, we thus rely on the following Monte Carlo approximation:
\begin{align}
\label{QFunction_approx}
Q (\bs{\phi}; \bs{\phi}^\star)  \approx  \tilde{Q} & (\bs{\phi}; \bs{\phi}^\star) \\
 \overset{c}{=}  - \frac{1}{R} \sum_{r=1}^{R} \sum_{(f,n)} & \bigg( \ln\left( g_n \sigma_f(\mathbf{z}_n^{(r)},\vb_n) + 
\left(\mathbf{W}_b\mathbf{H}_b\right)_{f,n} \right) \nonumber \\
& +  \frac{|x_{fn}|^2}{g_n \sigma_f(\mathbf{z}_n^{(r)},\vb_n) + \left(\mathbf{W}_b\mathbf{H}_b\right)_{f,n}} \bigg), \nonumber
\end{align}
where $\overset{c}{=}$ denotes equality up to additive terms that do not depend on $\bs{\phi}$ and $\bs{\phi}^\star$, and where
$\{\mathbf{z}_n^{(r)}\}_{r=1}^R$ is a sequence of samples drawn from the posterior $p(\mathbf{z}_n | \mathbf{x}_n,\vb_n 
;\bs{\phi}^\star)$ using \ac{MCMC} sampling. 
In practice we use the Metropolis-Hastings algorithm \cite{Robert:2005:MCS:1051451}, which forms the basis of the \ac{MCEM} algorithm 
\cite{wei1990monte}. At the $m$-th iteration of the Metropolis-Hastings algorithm and independently for all $n \in \{0,...,N-1\}$, a sample $\mathbf{z}_n$ is first drawn from a proposal random walk distribution:
\begin{equation}
\mathbf{z}_n | \mathbf{z}_n^{(m-1)} ; \epsilon^2 \sim \mathcal{N}(\mathbf{z}_n^{(m-1)}, \epsilon^2 \mathbf{I}).
\label{eq:mh1}
\end{equation}
Using the fact that this is a symmetric proposal distribution \cite{Robert:2005:MCS:1051451}, the acceptance probability 
$\eta$ is computed by: 
\begin{align*}
\eta =&\min\left(1, \frac{p( \mathbf{x}_n | \mathbf{z}_n,\vb_n ; \bs{\phi}^\star) p(\mathbf{z}_n|\vb_n;\gammab^\star)}{p( \mathbf{x}_n | 
\mathbf{z}_n^{(m-1)},\vb_n ; \bs{\phi}^\star) p(\mathbf{z}_n^{(m-1)}|\vb_n;\gammab^\star)}\right),
\label{log_acceptance_ratio}
\end{align*}
where
\begin{equation}
p\left( \mathbf{x}_n | \mathbf{z}_n,\vb_n ; \bs{\phi}^\star \right) = \prod_{f=0}^{F-1} p(x_{fn} | \mathbf{z}_n,\vb_n ; 
\bs{\theta}_{u}^\star),
\end{equation}
with $p(x_{fn} | \mathbf{z}_n,\vb_n ; \bs{\theta}_{u}^\star)$ defined in \eqref{likelihood} and $p\left(\mathbf{z}_n|\vb_n;\gammab^\star\right)$ 
defined in \eqref{eq:model_cvae}. 
Next, $u$ is drawn from a uniform distribution $\mathcal{U}([0,1])$. If $u < \eta$,  the sample is accepted and we set $\mathbf{z}_n^{(m)} = 
\mathbf{z}_n$, otherwise the sample is rejected and we set $\mathbf{z}_n^{(m)} = \mathbf{z}_n^{(m-1)}$. Only the last $R$ samples are kept for 
computing $\tilde{Q}(\bs{\phi}; \bs{\phi}^\star)$ in \eqref{QFunction_approx}, i.e. the samples drawn during a so called 
burn-in period are discarded.

\subsubsection{M-Step}

$\tilde{Q}(\bs{\phi}; \bs{\phi}^\star)$ in \eqref{QFunction_approx} is maximized with 
respect to the new model parameters $\bs{\phi}$. As usual in the NMF literature \cite{fevotte2011algorithms}, we adopt 
a block-coordinate approach by successively and individually updating $\mathbf{H}_b$, $\mathbf{W}_b$ and $\mathbf{g}$, using the auxiliary 
function technique as done in \cite{Leglaive_MLSP18}. Following the same methodology, we obtain the following formula for updating the NMF 
model parameters: 
\begin{equation}
\mathbf{H}_b \leftarrow \mathbf{H}_b \odot \left( \frac{\mathbf{W}_b^\top \left( | \mathbf{X} |^{\odot 2} \odot \sum\limits_{r=1}^{R} 
\left(\mathbf{V}_x^{(r)} \right)^{\odot-2} \right)}{\mathbf{W}_b^\top \sum\limits_{r=1}^{R} \left(\mathbf{V}_x^{(r)}\right)^{\odot-1} } 
\right)^{\odot 1/2},
\label{updateH}
\end{equation}
\begin{equation}
\mathbf{W}_b \leftarrow \mathbf{W}_b \odot \left( \frac{\left( | \mathbf{X} |^{\odot 2} \odot \sum\limits_{r=1}^{R} 
\left(\mathbf{V}_x^{(r)}\right)^{\odot-2} \right) \mathbf{H}_b^\top}{\sum\limits_{r=1}^{R} \left(\mathbf{V}_x^{(r)}\right)^{\odot-1} 
\mathbf{H}_b^\top } \right)^{\odot 1/2},
\label{updateW}
\end{equation}
where $(\cdot)^{\odot (\cdot)}$ denotes element-wise exponentiation, $(\cdot)\odot(\cdot)$ denotes element-wise multiplication, and $\frac{(\cdot)}{(\cdot)}$
denotes element-wise division. Moreover, $\mathbf{V}_x^{(r)} \in 
\mathbb{R}_+^{F \times N}$ is the matrix with entries $g_n \sigma_f(\mathbf{z}_n^{(r)},\vb_n) + 
\left(\mathbf{W}_b\mathbf{H}_b\right)_{f,n}$, and $\mathbf{X} \in \mathbb{C}^{F \times N}$ is the matrix with entries $ 
\left(\mathbf{X}\right)_{f,n} = x_{fn}$. 
The gains are updated as follows:
\begin{equation}
\mathbf{g}^\top \leftarrow \mathbf{g}^\top \odot \left( \frac{ \mathbf{1}^\top  \left(| \mathbf{X} |^{\odot 2} \odot \sum\limits_{r=1}^{R} 
\left(\mathbf{V}_s^{(r)} \odot \left(\mathbf{V}_x^{(r)} \right)^{\odot-2}\right)\right)}{\mathbf{1}^\top \left[ \sum\limits_{r=1}^{R} 
\left(\mathbf{V}_s^{(r)} \odot \left(\mathbf{V}_x^{(r)} \right)^{\odot-1}\right)\right]} \right)^{\odot 1/2},
\label{update_g}
\end{equation}
where $\mathbf{1}$ is a vector of ones of dimension $F$ and $\mathbf{V}_s^{(r)} \in \mathbb{R}_+^{F \times N}$ is the matrix with 
entries $\sigma_f(\mathbf{z}_n^{(r)},\vb_n) $. The nonnegative property of $\mathbf{H}_b$, $\mathbf{W}_b$ and of $\mathbf{g}$ is 
ensured, provided that their entries are initialized with nonnegative values. In practice, only one 
iteration of updates \eqref{updateH}, \eqref{updateW} and \eqref{update_g} is performed at each M-step.
\begin{algorithm}[t]
	\caption{Audio-visual CVAE speech enhancement}\label{alg:prop}
	\begin{algorithmic}[1]
		\State \textbf{Inputs:} \begin{itemize}
			\item Learned CVAE generative model for clean speech, i.e., \eqref{eq:model_cvae} and \eqref{encoder_CVAE}
			\item Noisy microphone frames $ \mathbf{x} = \lk 
			\mathbf{x}_n\rk_{n=0}^{N-1} $
			\item Video frames $ \mathbf{v} = \lk \mathbf{v}_n\rk_{n=0}^{N-1} $
		\end{itemize} 
		\State \textbf{Initialization:}
		\begin{itemize}
			\item Initialization of NMF noise parameters $ \Hb_b $ and $ \Wb_b $ with random nonnegative values
			\item Initialization of latent codes $ \mathbf{z} = \lk \mathbf{z}_n\rk_{n=0}^{N-1} $ using the learned encoder network \eqref{encoder_CVAE}  with $ \mathbf{x} = \lk 
			\mathbf{x}_n\rk_{n=0}^{N-1} $ and $ \mathbf{v} = \lk \mathbf{v}_n\rk_{n=0}^{N-1} $
			\item Initialization of the gain vector $\mathbf{g} = (g_0 \dots g_{N-1})^\top = \mathbf{1}$
		\end{itemize} 	
		\While{stop criterion not met}:
		\State{ \textbf{E-step:} Compute \eqref{QFunction_approx} using the Metropolis-Hastings \hspace*{1.75cm} algorithm}
		\State \textbf{M-$ \Hb_b $-step:} Update $ \Hb_b $ using \eqref{updateH}
		\State \textbf{M-$ \Wb_b $-step:} Update $ \Wb_b $ using \eqref{updateW}
		\State \textbf{M-$ \mathbf{g} $-step:} Update $ \mathbf{g} $ using \eqref{update_g}
		\EndWhile
		\State \textbf{Speech reconstruction:} Estimate $ \mathbf{s} = \lk 
		\mathbf{s}_n\rk_{n=0}^{N-1} $ with \eqref{source_estimate}
		
	\end{algorithmic}
	\vspace{-2pt}
	\label{algo:metropolis-within-gibbs}
\end{algorithm}

\subsection{Speech Reconstruction}
\label{subsec:speech_rec}

Let $\bs{\phi}^\ast = \{\mathbf{W}_b^\ast, \mathbf{H}_b^\ast, \mathbf{g}^\ast\}$ denote the set of parameters estimated by the above 
MCEM algorithm. Let $\tilde{s}_{fn} = \sqrt{g_n^\ast} s_{fn}$ be the scaled version of the speech STFT 
coefficients as introduced in \eqref{mixture}, with $g_n^\ast = (\mathbf{g}^\ast)_n$. The final step is to estimate these coefficients 
according to their posterior mean \cite{Leglaive_MLSP18}:
\begin{align}
\label{source_estimate}
\hat{\tilde{s}}_{fn} &= \mathbb{E}_{p(\tilde{s}_{fn} | x_{fn},\vb_n ; \bs{\phi}^\ast)}  [\tilde{s}_{fn}]  \\
&= \mathbb{E}_{p(\mathbf{z}_n | \mathbf{x}_n,\vb_n ; \bs{\phi}^\ast) }\left[ \mathbb{E}_{p(\tilde{s}_{fn} | \mathbf{z}_n,\vb_n, 
\mathbf{x}_n ; \bs{\phi}^\ast)} [\tilde{s}_{fn}] \right]\nonumber \\
&= \mathbb{E}_{p(\mathbf{z}_n | \mathbf{x}_n,\vb_n ; \bs{\phi}^\ast) }\left[\frac{g_n^\ast \sigma_f^2(\mathbf{z}_n,\vb_n)}{g_n^\ast 
\sigma_f^2(\mathbf{z}_n,\vb_n) + (\mathbf{W}_b^\ast\mathbf{H}_b^\ast)_{f,n}}\right]x_{fn}. \nonumber
\end{align}
This estimation corresponds to a ``probabilistic'' version of Wiener filtering, with an averaging of the filter over the posterior distribution of the latent variables. 
As above, this expectation cannot be computed analytically, but instead it can be approximated using the same Metropolis-Hastings algorithm of Section~\ref{subsec:EStep}. The time-domain estimate of the speech signal is finally obtained from the inverse STFT with overlap-add.

The complete speech enhancement procedure is summarized in Algorithm~\ref{alg:prop}, which we refer to as AV-CVAE speech enhancement.
\section{Implementation and Experiments}\label{sec:exp}

\subsection{The NTCD-TIMIT Dataset}
\label{subsec:NTCT-TIMIT}

 \addnote[exp-1]{1}{For training, validation and test we used the NTCD-TIMIT dataset \cite{Abde17}, which contains \ac{AV} recordings from $ 56 $ English speakers with an Irish
accent, uttering $98$ different sentences. So there are $56\times98 =5488 $ videos with an approximate length of $5$ seconds \cite{TIMIT}. The visual data consists of 30~FPS videos of lip \acp{ROI}. Each frame (\ac{ROI}) is of size 67$\times$67 pixels. The speech signal is sampled at $16$ kHz. The audio spectral features are computed using an STFT window of 64~ms ($1024$ samples per frame) with 47.9\% overlap, hence $F=513$. This guarantees that the audio frame rate is equal to the visual frame rate and that the frames associated with the two modalities are synchronized. }

 In addition to clean signals, noisy versions are also provided, with six types of noise, namely \textit{\ac{LR}}, \textit{White}, \textit{Cafe}, \textit{Car}, \textit{Babble}, and \textit{Street}. For each noise type, there are five noise levels: $-5$~dB, $0$~dB, $5$~dB, $15$~dB and $20$~dB. In all the experiments, we used five sequences per speaker, and six noise levels, $-15$~dB, $ -10 $~dB, $-5$~dB, $ 0 $~dB, $5$~dB, and $15$~dBs to test the performance of different architectures.\footnote{Note that the recordings with noise levels of $ -15 $~dB and $ -10 $~dB are not provided with the NTCD-TIMIT dataset. Hence, we created noisy versions by following the same procedure as in \cite{Abde17}, which is based on the FaNT filtering and noise-adding tool \cite{Hirs005}.} The dataset is divided into $39$ speakers for training, $8$ speakers for validation, and $9$ speakers for testing, as proposed in \cite{Abde17}. The speakers in each split are different. The length of clean-speech corpora used for training is of about $ 5 $ hours, while the length of noisy speech used for testing is of $ 1 $ hour.

\subsection{The GRID Dataset}
\label{subsec:GRID}
\addnote[exp-grid]{1}{For the test phase, in addition to the NTCD-TIMIT test set, we also experimented with the GRID dataset \cite{CookBCS06_grid}. This dataset contains clean \ac{AV} recordings of 34 people (16 women and 18 men), each uttering 1000 sentences. The duration of each utterance is about 3 seconds. We randomly selected 270 samples from this dataset as our test data. Because there is no noisy version for the audio recordings in this dataset, we created our own noisy test set by adding to each test sample different noise types with different noise levels. The noise types are the same as those in the NTCD-TIMIT dataset.}

\subsection{Model and Architecture Variants} 
\label{sec:vae-variants}

In order to assess the performance of the proposed \ac{AV} speech enhancement method and, in particular, to quantify the contribution of visual information, we implemented and tested the AV-CVAE architecture as well as several variants, namely
A-VAE \cite{Leglaive_MLSP18}, i.e. Section~\ref{sec:a_vae} and Fig.~\ref{fig:vae}, V-VAE, i.e. Section~\ref{sec:visual-vae} and Fig.~\ref{fig:v_vae}, and AV-VAE, i.e. a simplified version of AV-CVAE where the prior for $ \zb $ is a standard Gaussian distribution. As it can be observed in Fig.~\ref{fig:vae_encdec}, AV-CVAE combines A-VAE and V-VAE, we therefore describe in detail these two architectures.

\begin{figure*}[t!]
	\centering
	{\includegraphics[width=0.99\textwidth]{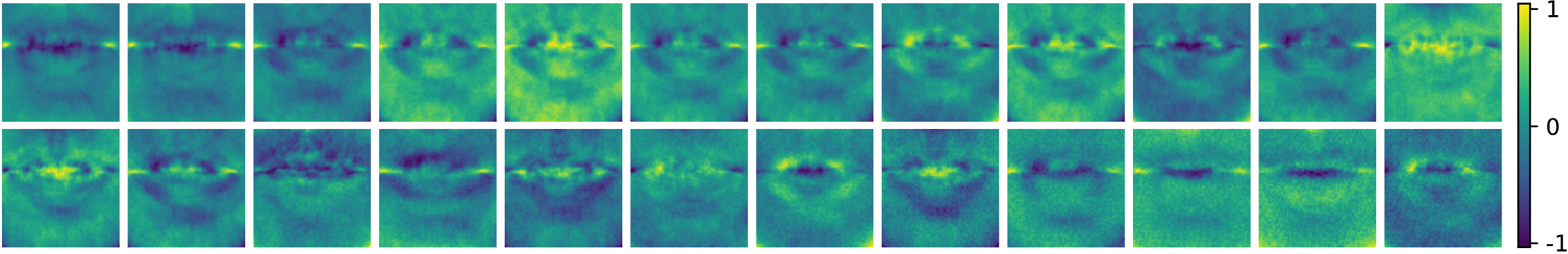}}
	\caption{\label{fig:weights} Each one of these 67$\times$67 images visualizes the learned values of the weights associated with a node of the first fully-connected layer of the base V-VAE architecture. This first layer contains 512 nodes, hence there is an equal number of images like the ones displayed here. This illustrates the effectiveness of the proposed model to extract speech features from lips.}
\end{figure*}

 A-VAE uses the same architecture as the one described in \cite{Leglaive_MLSP18}: both encoder and decoder have a single hidden layer with 128 nodes and hyperbolic tangent activations, \addnote[arch1]{2}{consisting of only fully-connected layers. The input dimension of the encoder is $ F=513 $. } The dimension of the latent space is $L=32$. \addnote[arch2]{2}{As already mentioned in Section~\ref{sec:visual-vae} and illustrated in Fig.~\ref{fig:v_vae}, V-VAE shares the same decoder architecture as A-VAE. The encoder is, however, different because of the need of additional visual feature extraction, as described below.}

We adopt two architectures for embedding lip \acp{ROI} into a feature vector $\mathbf{v} \in \mathbb{R}^M$, with $M=128$.
The first architecture, \textit{base} V-VAE, is composed of two fully-connected layers with 512 and 128 nodes, respectively. The dimension of the input corresponds to a single frame that is vectorized, namely $4489=67\times 67$. \addnote[arch3]{2}{A fully-connected layer then takes as input the feature vector of dimension $ M=128 $ and predicts as output the mean and variance parameters in the latent space}. The second architecture consists of the base network just described and augmented with a \textit{front-end} network composed of a 3D convolution layer, that takes as input five consecutive frames and which is constituted of 64 kernels of size $5\times 7\times 7$, followed by batch normalization and rectified linear units, and by a 34 layer ResNet, yielding an output of dimension $1280$ which is then passed to the base network. This front-end network is shown in a dashed box on Fig.~\ref{fig:v_vae}. The front-end network was trained as part of an \ac{AV} speech recognition deep architecture \cite{PetrSMC18} using a 500 word vocabulary.

\addnote[arch4]{2}{The AV-CVAE architecture is similar to that of A-VAE, except for the fact that the visual features (extracted using one of the two networks described above) are concatenated with the speech-spectrogram time frames at the input of the encoder, as well as with the latent code at the input of the decoder}. Note that the visual network just described appears three times, once as part of the prior network for $ \zb $ and twice as part of AV-CVAE, hence two training strategies are possible, (i)~to constrain these three networks to share the same set of weights, or (ii)~to allow a different set of weights for each one of these networks. Through our experiments, we noticed that the former strategy yields better performance that the latter, which also has the advantages of dealing with fewer weights, of a lower chance of overfitting and of a reduced computational cost.

It is worth looking at the learned values of the weights associated with the first fully-connected layer of the base V-VAE architecture. In this case, the input consists of $67\times 67$ lip \acp{ROI} which is fed into a layer with 512 nodes. Hence, there are $67\times 67$ weights per node. \addnote[exp-visu1]{1}{The trained weights of the first layer of the visual sub-network are multiplied (in an element-wise way) with the input, namely raw lip images. As such, these weights extract important features present in the lip images. These weights give us some insights about how the network exploits the visual data. } One can visualize these nodes as an image, e.g. Fig.~\ref{fig:weights} in which 24 such images are displayed. This illustrates the effectiveness of the proposed V-VAE model to extract salient visual speech features. \addnote[exp-visu2]{1}{It can be seen that the shown weights exhibit lip-like shapes. The parts of the weight images with higher intensities correspond to regions on the input lip images that are relevant for the task at hand.}

\subsection{State of the Art Method}

The \ac{AV} speech enhancement method we propose is unsupervised, i.e. it does \emph{not} require pairs of clean and noisy speech signals for training. Nevertheless, it is interesting to compare its performance with a supervised method. We compare our approach with the recently proposed state-of-the-art supervised method \cite{GabbSP18}, which is close to the one proposed in \cite{hou2018audio}, and whose Python implementation is publicly available online.\footnote{\url{https://github.com/avivga/audio-visual-speech-enhancement}} This method is based on a \ac{DNN} made of two subnetworks, one for processing spectrograms of noisy speech and another one for processing lip \acp{ROI}. The resulting audio and visual
encodings are then concatenated and processed to yield a single embedding which is fed into a network with three fully-connected layers. 
Finally, a spectrogram of the enhanced speech is obtained using an audio decoder \cite{GabbSP18}. 

%
As is the case with the front-end visual network described above \cite{PetrSMC18}, 128$\times$128 frames containing lip \acp{ROI} are extracted from the raw video associated with a speaker and the visual input to the network is composed of five consecutive frames. The audio input is a spectrogram synchronized with these five frames. 
For training, we used the parameter setting suggested by the authors of \cite{GabbSP18}. Training this supervised method requires noisy mixtures. We used the DEMAND dataset \cite{DEMAND_db} to add various noise types to the clean-speech sequences of the 
NTCD-TIMIT dataset, with various SNR levels. We used the same SNR levels as in the NTCDT-TIMIT test dataset. The noise types of the DEMAND dataset are different than the ones that were used to generate noisy-speech instances described in Section~\ref{subsec:NTCT-TIMIT}, although they share similarities.

\addnote[exp-nmf]{1}{In addition to the supervised audio-visual method described above, we also compare the performance of the proposed method with an audio-only \ac{NMF} baseline. It corresponds to a framework referred to as ``semi-supervised source separation'' in the literature \cite{SmarRS07_nmf, MysoS11, mohammadiha2013supervised}, where during a training phase, we first learn a speech \ac{NMF} dictionary from a dataset of clean speech signals. Then, at test time, given the learned dictionary, the speech activation matrix and the noise NMF model parameters are estimated from the noisy mixture signals. The speech signal is then estimated by Wiener filtering.}
\subsection{Implementation Details and Parameter Settings}

As mentioned above, VAE training requires a gradient-ascent method. In practice we used the Adam optimizer \cite{kingma2014adam} with a step size of $10^{-4}$. Using the NTCD-TIMIT validation set, early stopping was used with a patience of $ 20 $ epochs. To alleviate the effect of random initialization, we have trained each VAE model five times. The performance was measured by averaging across the sequences present in the dataset (see below) as well as across these five trained models.

The rank of the \ac{NMF} noise model is set to $K=10$ and the parameters of the nonnegative matrices associated with this model are randomly initialized with nonnegative values. For the baseline NMF method \cite{SmarRS07_nmf}, the rank of the NMF speech model is set to $ 64 $. Similarly to \cite{bando2018statistical,Leglaive_MLSP18}, at the first iteration of the MCEM algorithm, the Markov chain of the Metropolis-Hastings algorithm was initialized using the mixture signal and the corresponding visual features as input to the encoder. That is, for all $l \in \{0,...,L-1\}$, $z_{ln}^{(0)} = \tilde{\mu}_l\left(\tilde{\mathbf{x}}_n, \vb_n\right)$, where we used the same notation as above, namely 
$ \tilde{\mathbf{x}}_n \triangleq (|x_{0n}|^2 \dots |x_{F-1\: n}|^2)^{\top}$

\begin{figure}[t!]
	\centering
	\subfloat[PESQ]{{\includegraphics[height=4cm]{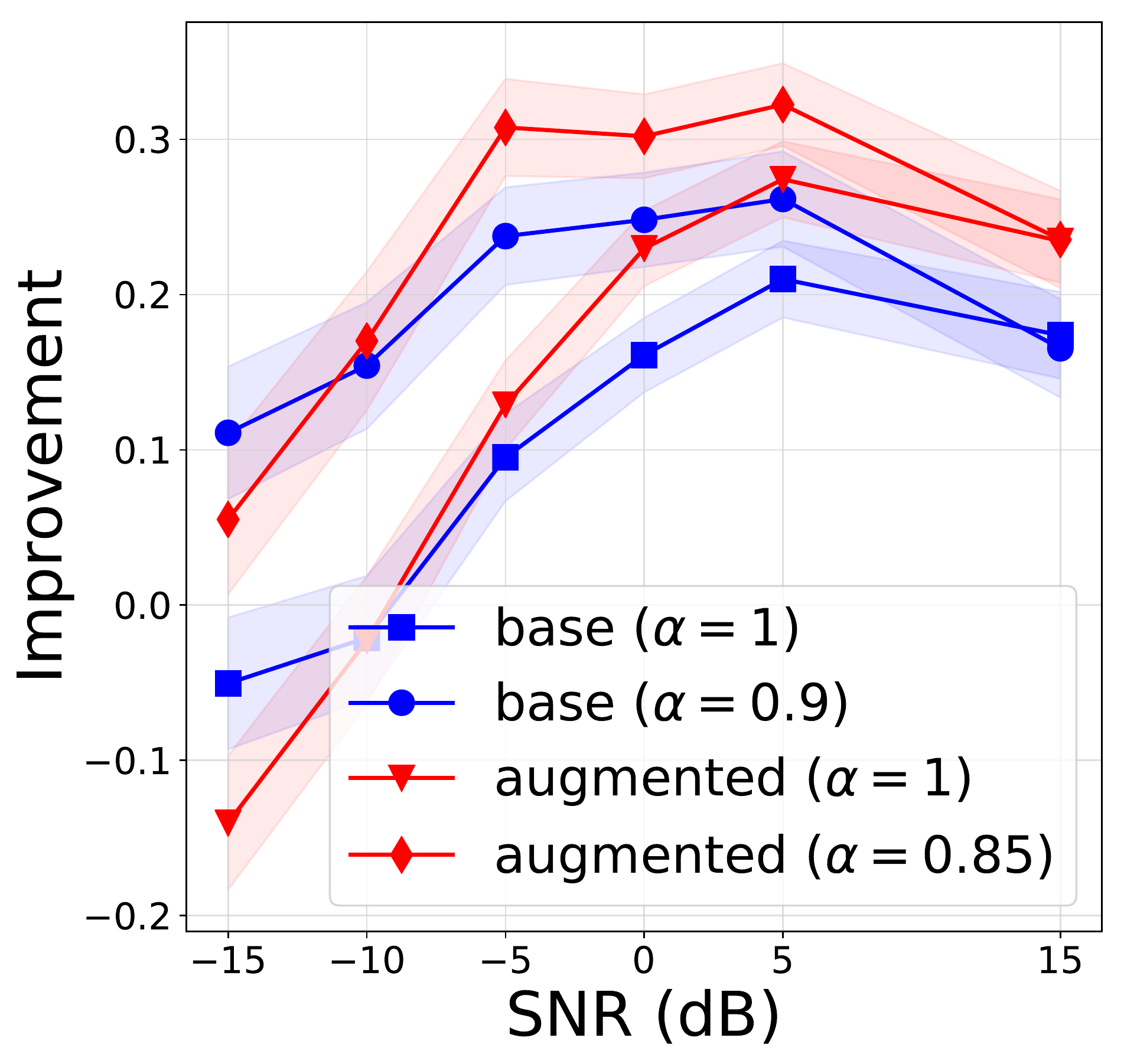} }}
	\subfloat[SDR]{{\includegraphics[height=4cm]{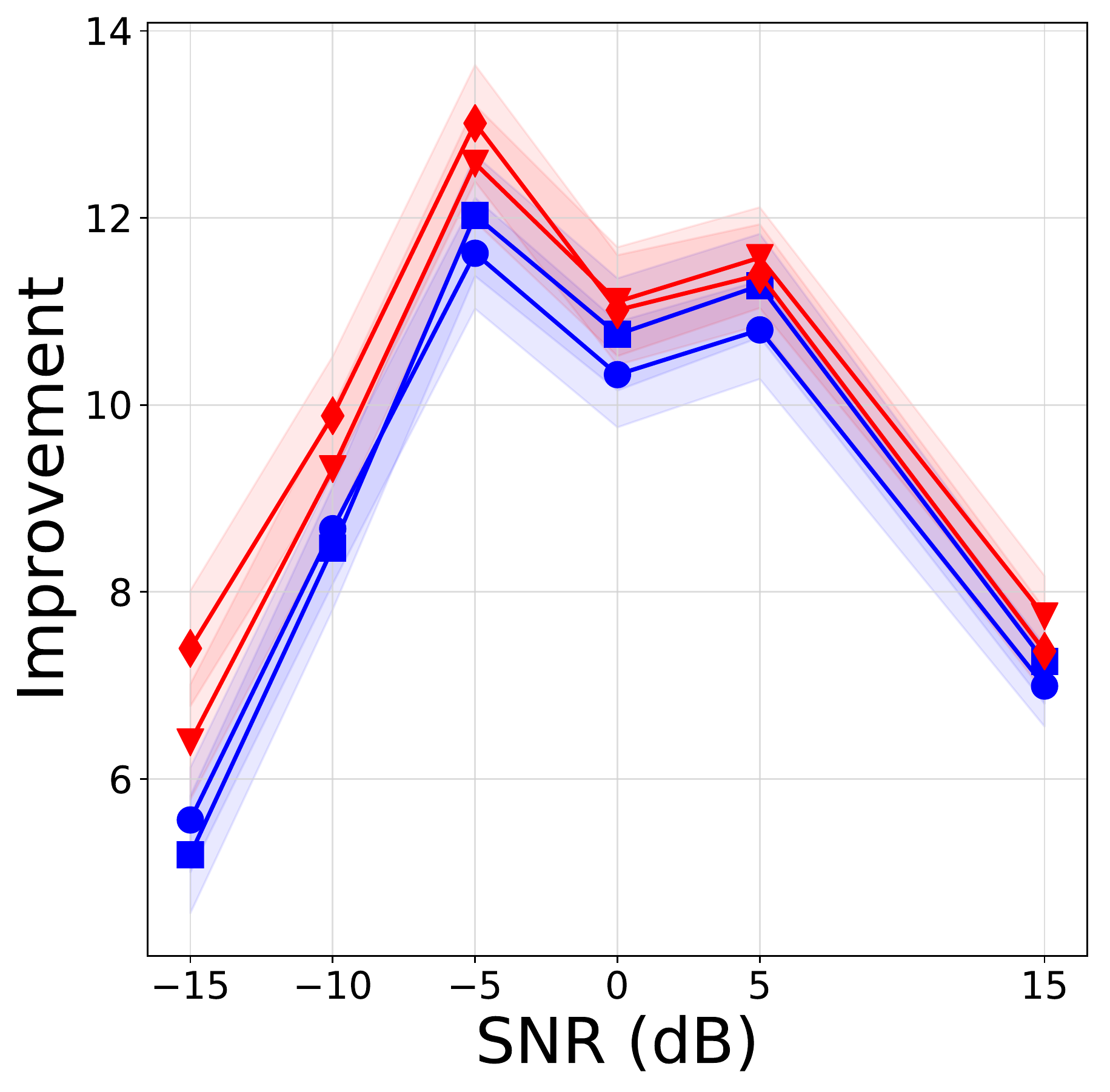} }}
	\caption{\label{fig:cvae_bcvae} Performance of AV-CVAE  for two different values of $ \alpha $ in \eqref{eq:cvae_elbotild}, and for the two variants of visual feature embedding.}
\end{figure}

\begin{figure}[t!]
	\centering
	\subfloat[PESQ]{{\includegraphics[height=4cm]{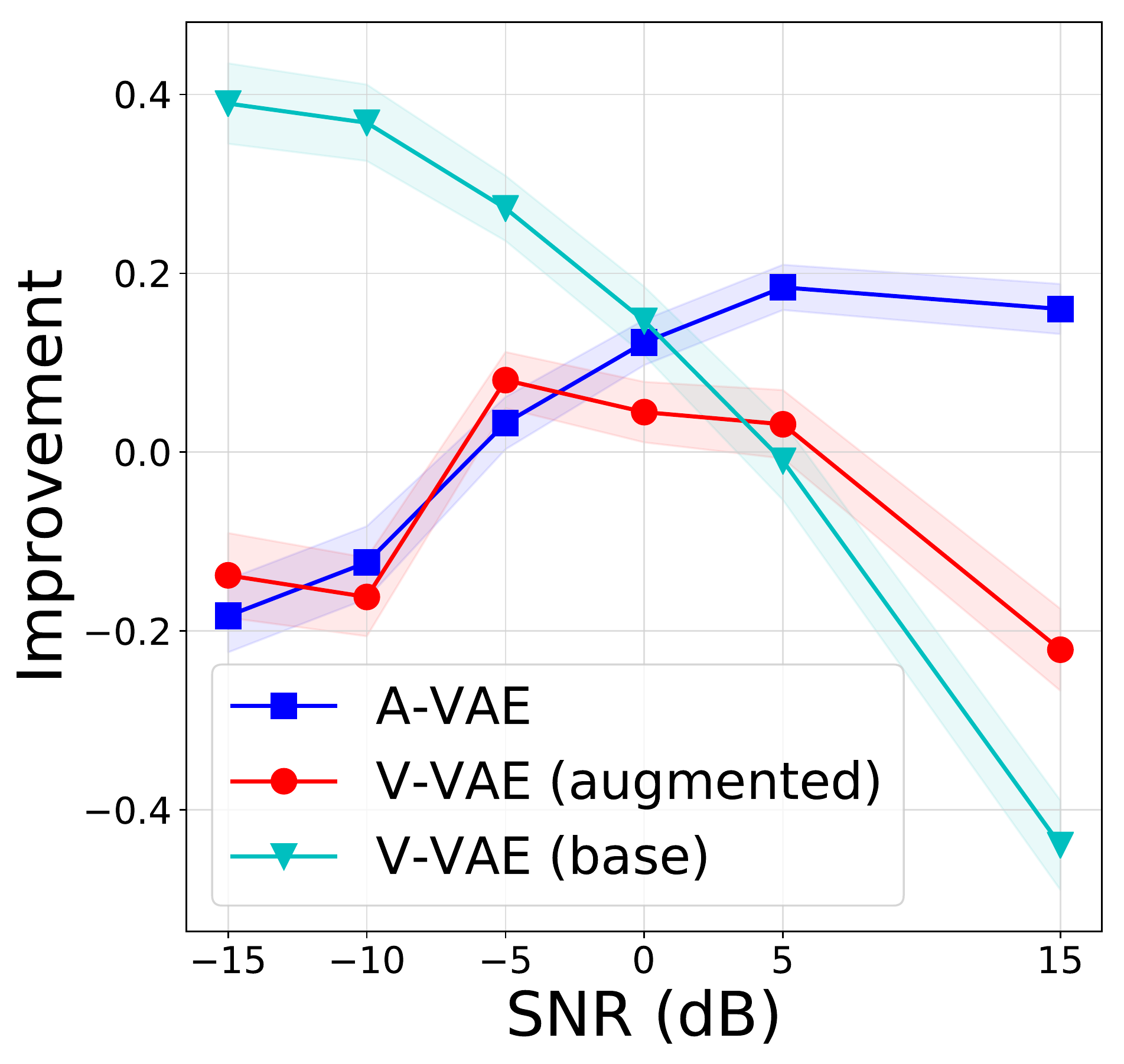} }}
	\subfloat[SDR]{{\includegraphics[height=4cm]{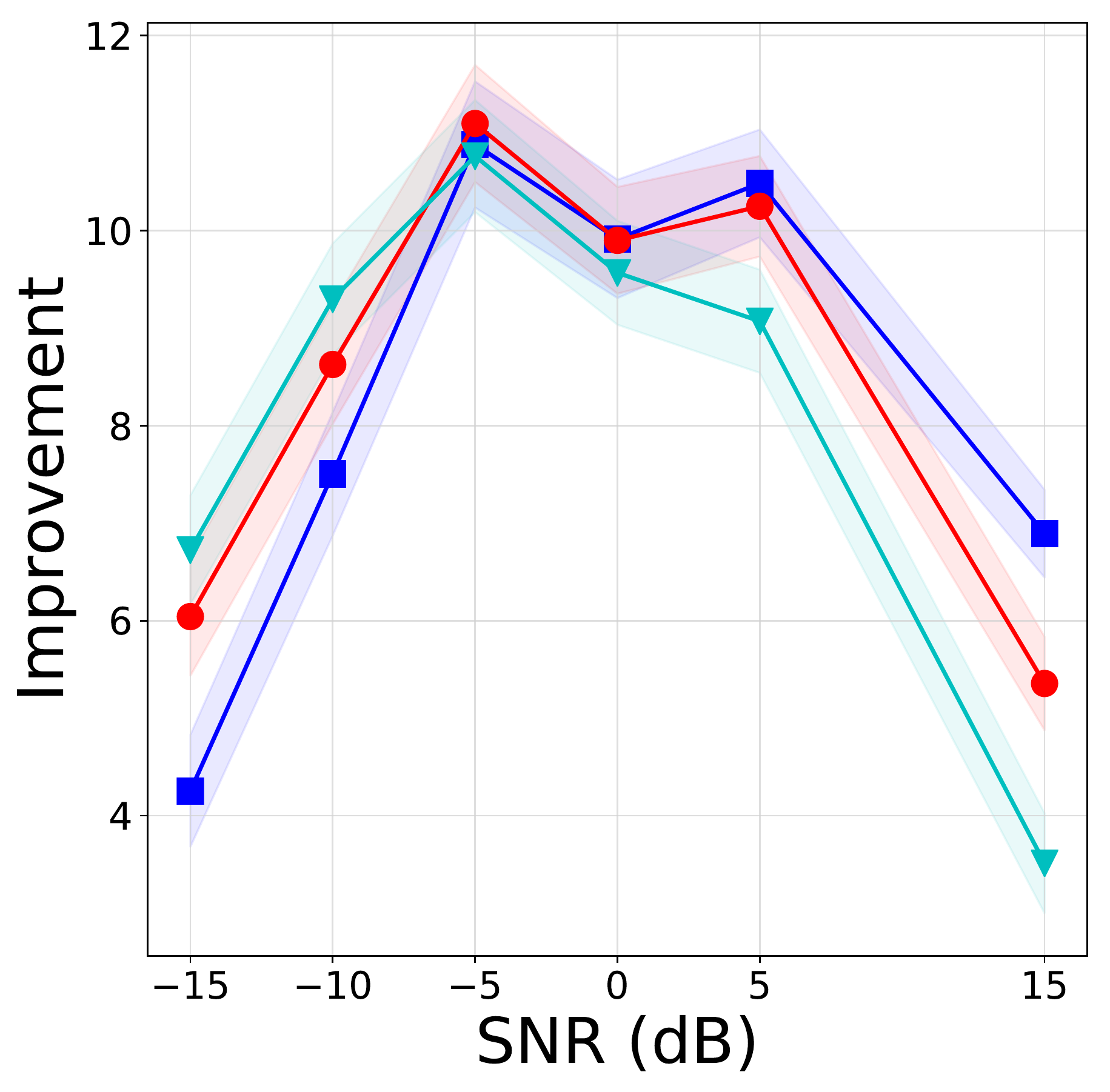} }}
	\caption{\label{fig:v_a} Performance comparison of A-VAE and of the two variants of V-VAE.}
\end{figure}

\begin{figure}[t!]
	\centering
	\subfloat[PESQ]{{\includegraphics[height=4cm]{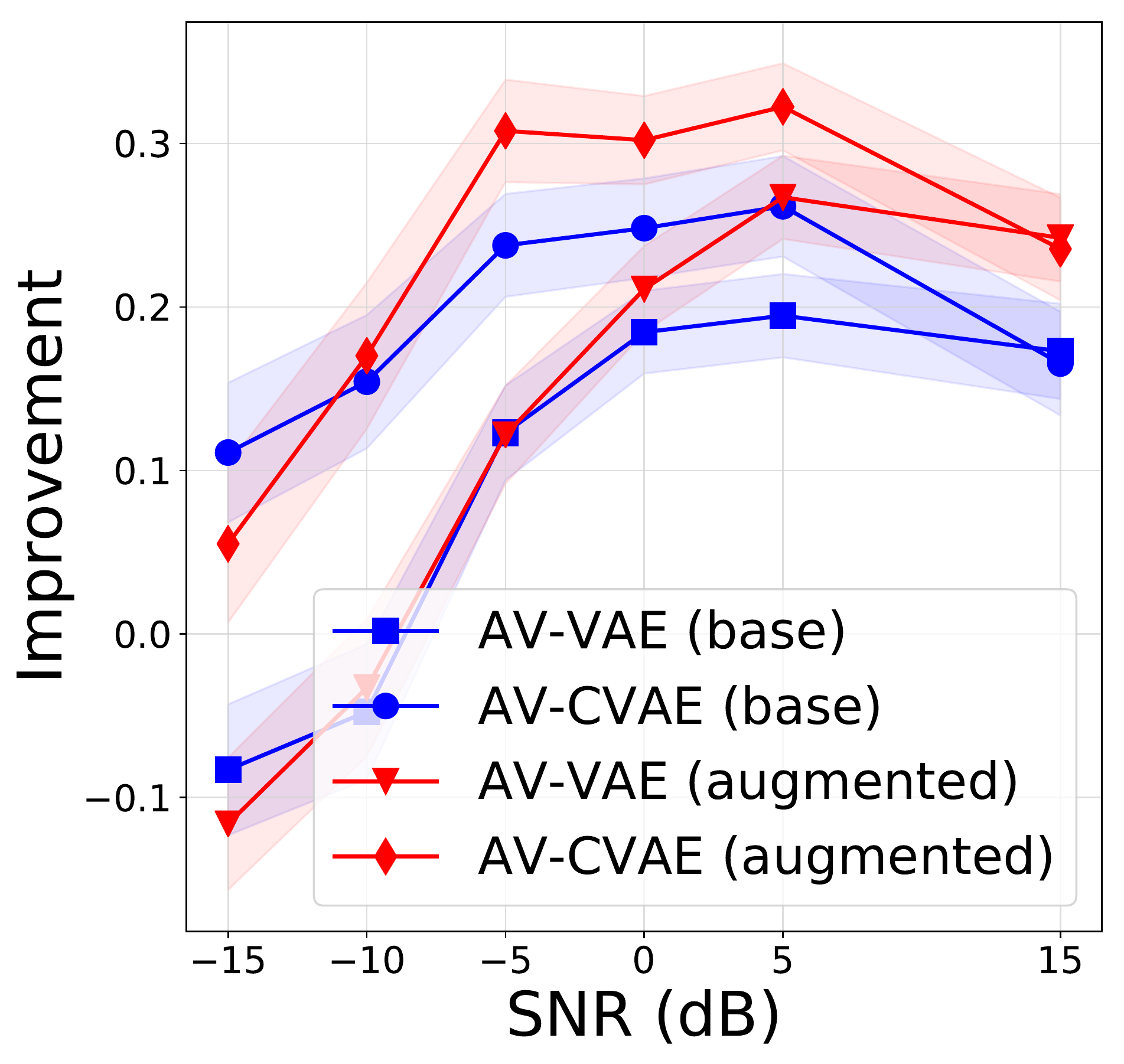} }}
	\subfloat[SDR]{{\includegraphics[height=4cm]{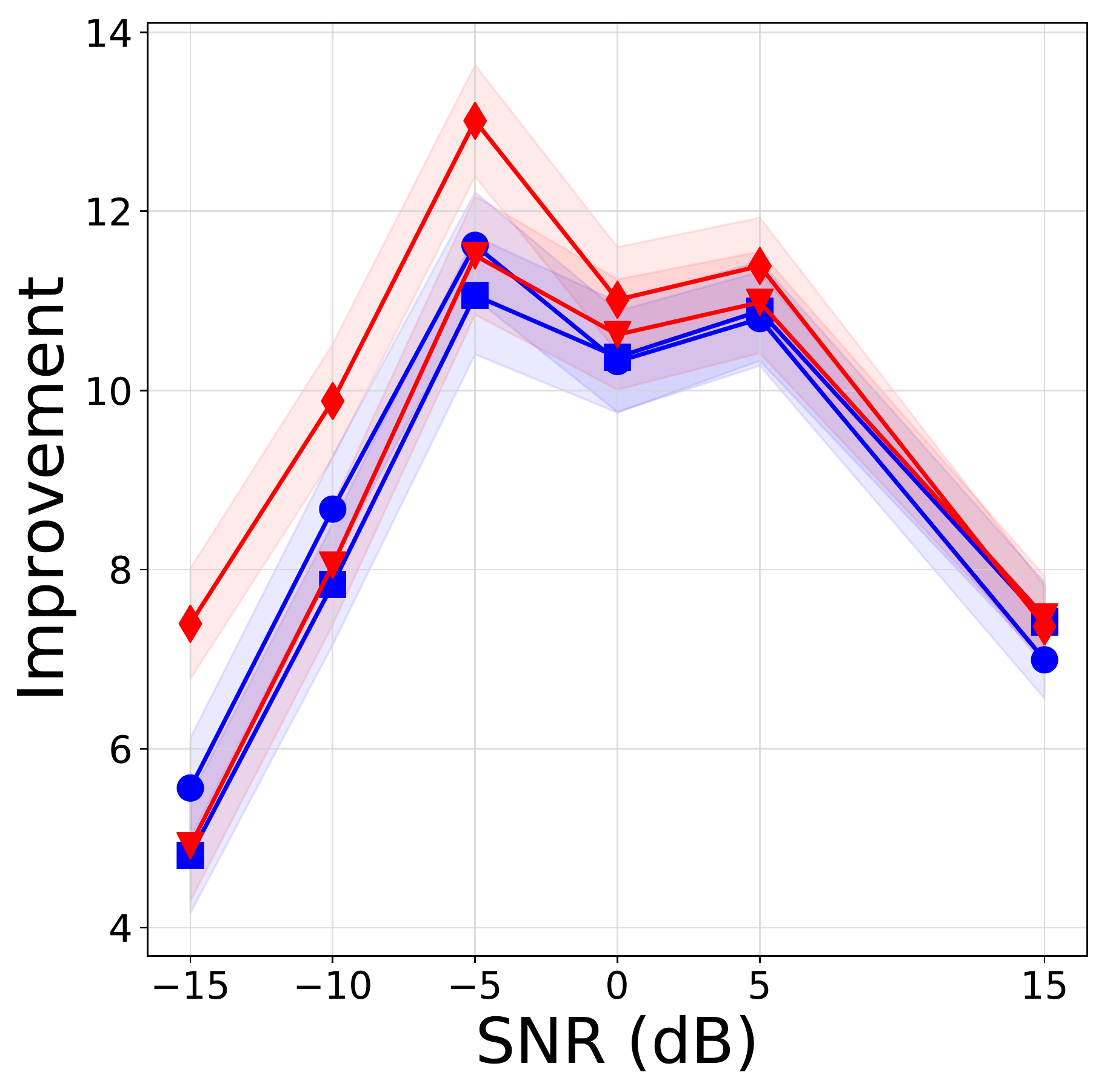} }}
	\caption{\label{fig:vae_cvae} Performance and comparison of AV-VAE and AV-CVAE.}
\end{figure}

\subsection{Results}

We used standard speech enhancement scores, namely the \ac{SDR} \cite{vincent2006performance}, the \ac{PESQ}  \cite{rix2001perceptual}, \addnote[exp-stoi]{1}{and the \ac{STOI} scores \cite{taal2011algorithm}}. \ac{SDR} is measured in decibels (dB), while \ac{PESQ} and  \ac{STOI} values lie in the intervals $[-0.5,4.5]$ and $[0,1]$, respectively (the higher the better). For computing SDR the mir\_eval Python library was used.\footnote{\url{https://github.com/craffel/mir_eval}} For each measure, we report the difference between the output value, i.e., evaluated on the enhanced speech signal, and the input value, i.e., evaluated on the noisy/unprocessed mixture signal.

We compared the proposed unsupervised AV-CVAE method and its variants with the supervised state of the art method outlined above,  \cite{GabbSP18}, \addnote[exp-nmf]{1}{and the unsupervised \ac{NMF} method inspired from \cite{SmarRS07_nmf}}. For each experiment, the median values of SDR, PESQ, and STOI scores, along with their corresponding standard errors, computed over all noise types, all test samples, and over five models obtained with five random initializations. The SDR, PESQ, and STOI scores are plotted as a function of noise levels.

\begin{figure}[t]
	\centering
	\subfloat[PESQ]{{\includegraphics[height=4cm]{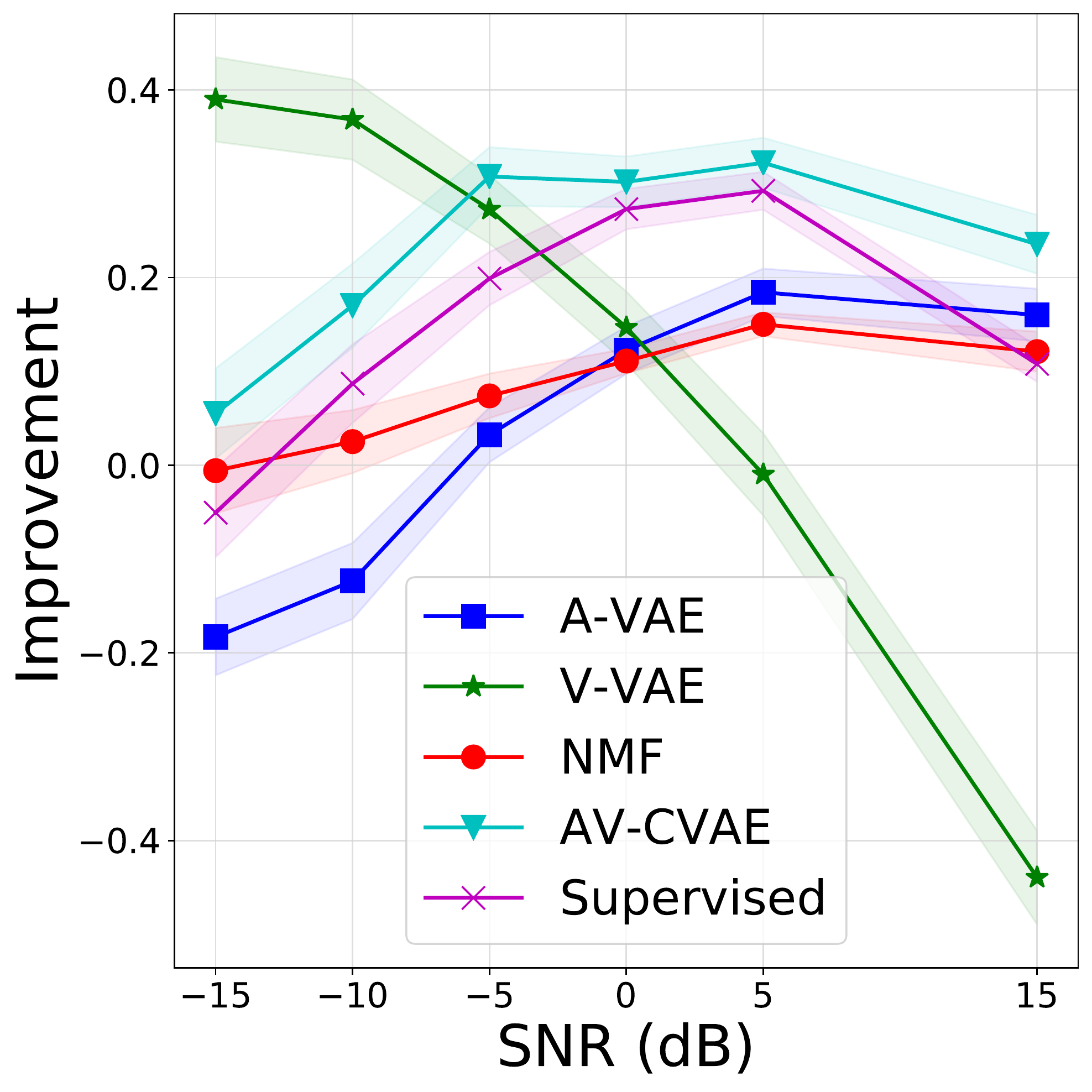} }}
	\subfloat[SDR]{{\includegraphics[height=4cm]{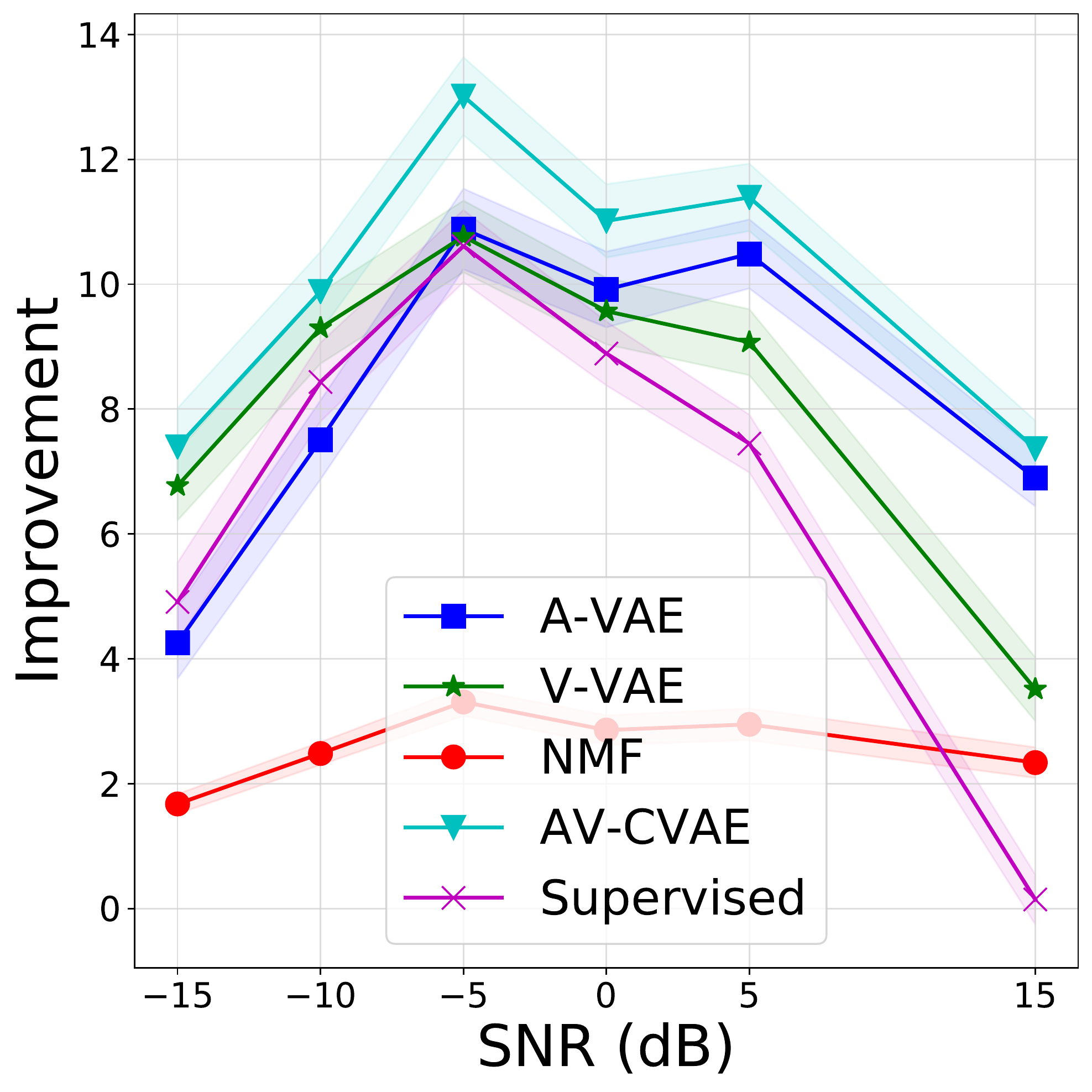} }}\\
	\subfloat[STOI]{{\includegraphics[height=4cm]{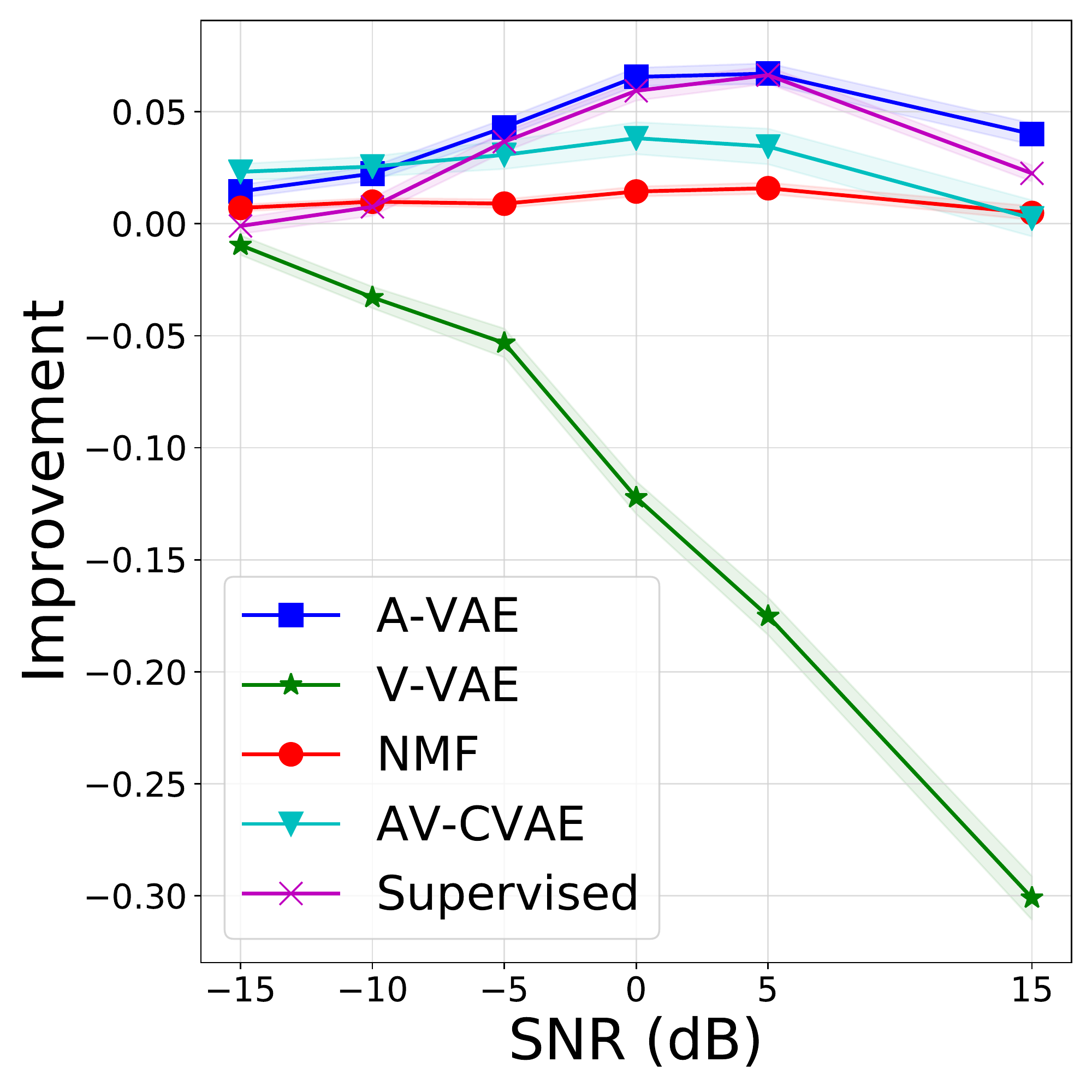} }}
	\caption{\label{fig:all} Performance comparison of A-VAE, V-VAE, NMF, AV-CVAE, and \cite{GabbSP18} on the test samples from the NTCD-TIMIT dataset.}
\end{figure}

\begin{figure}[t!]
	\centering
	\subfloat[LR]{{\includegraphics[height=4cm]{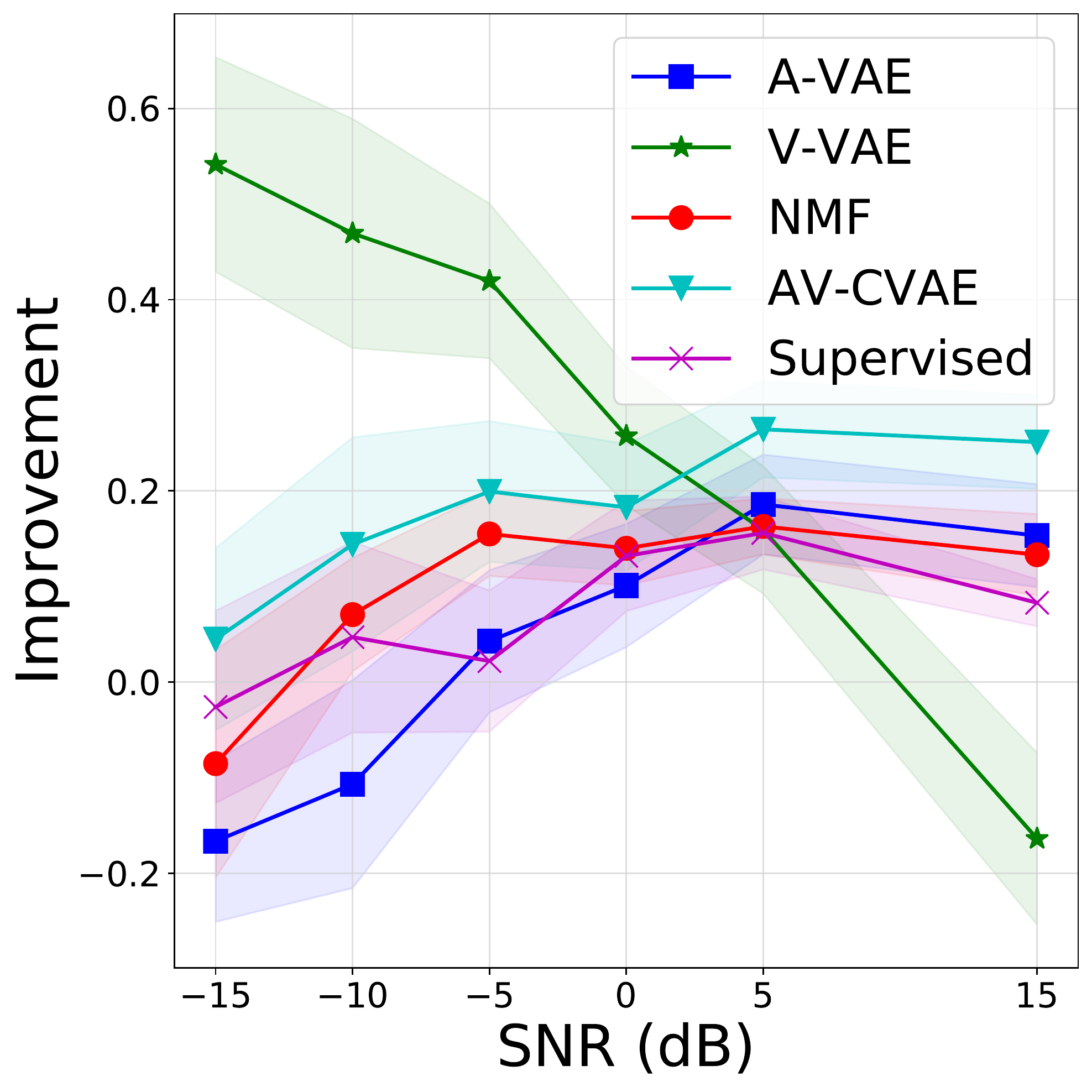} }}
	\subfloat[White]{{\includegraphics[height=4cm]{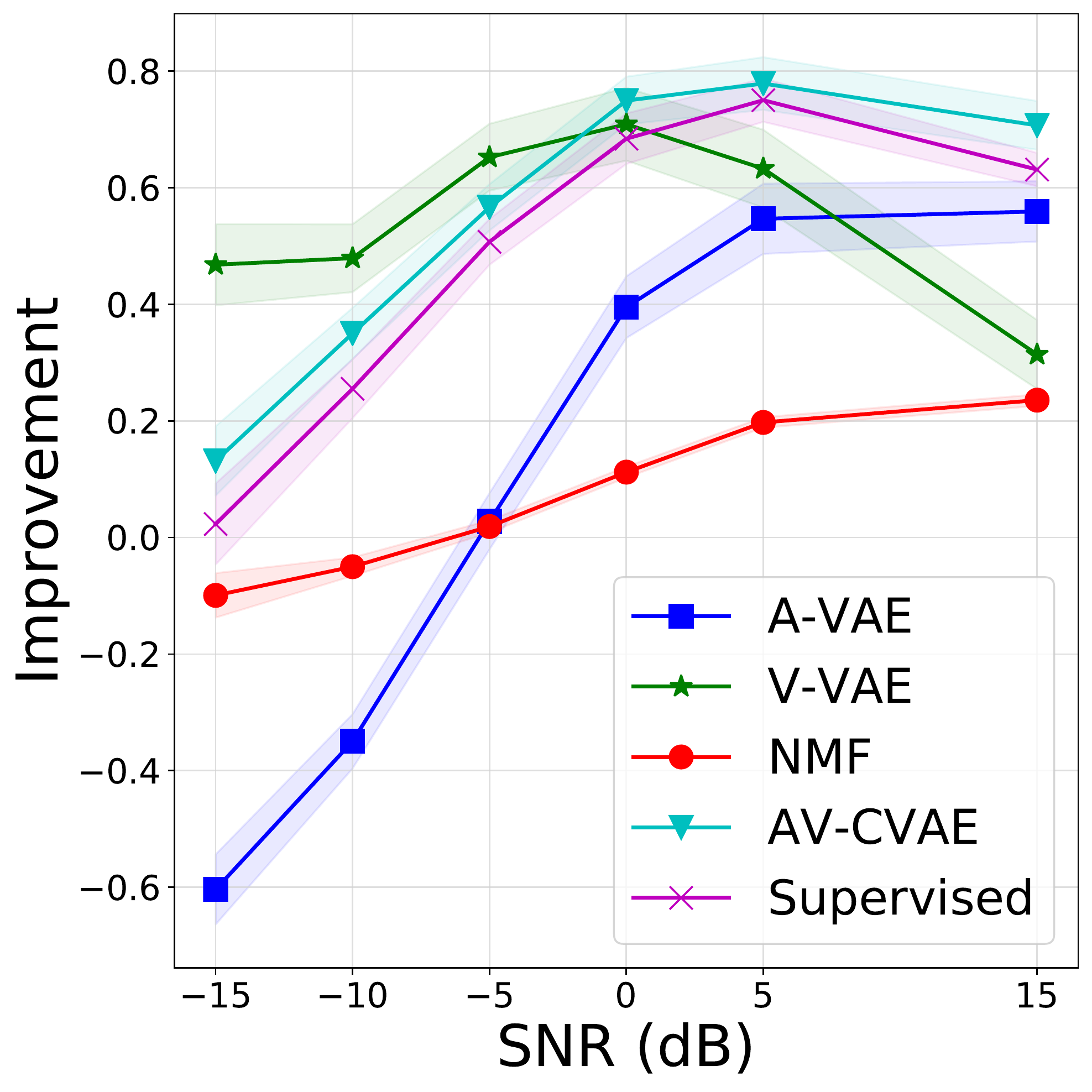} }}\\
	\subfloat[Street]{{\includegraphics[height=4cm]{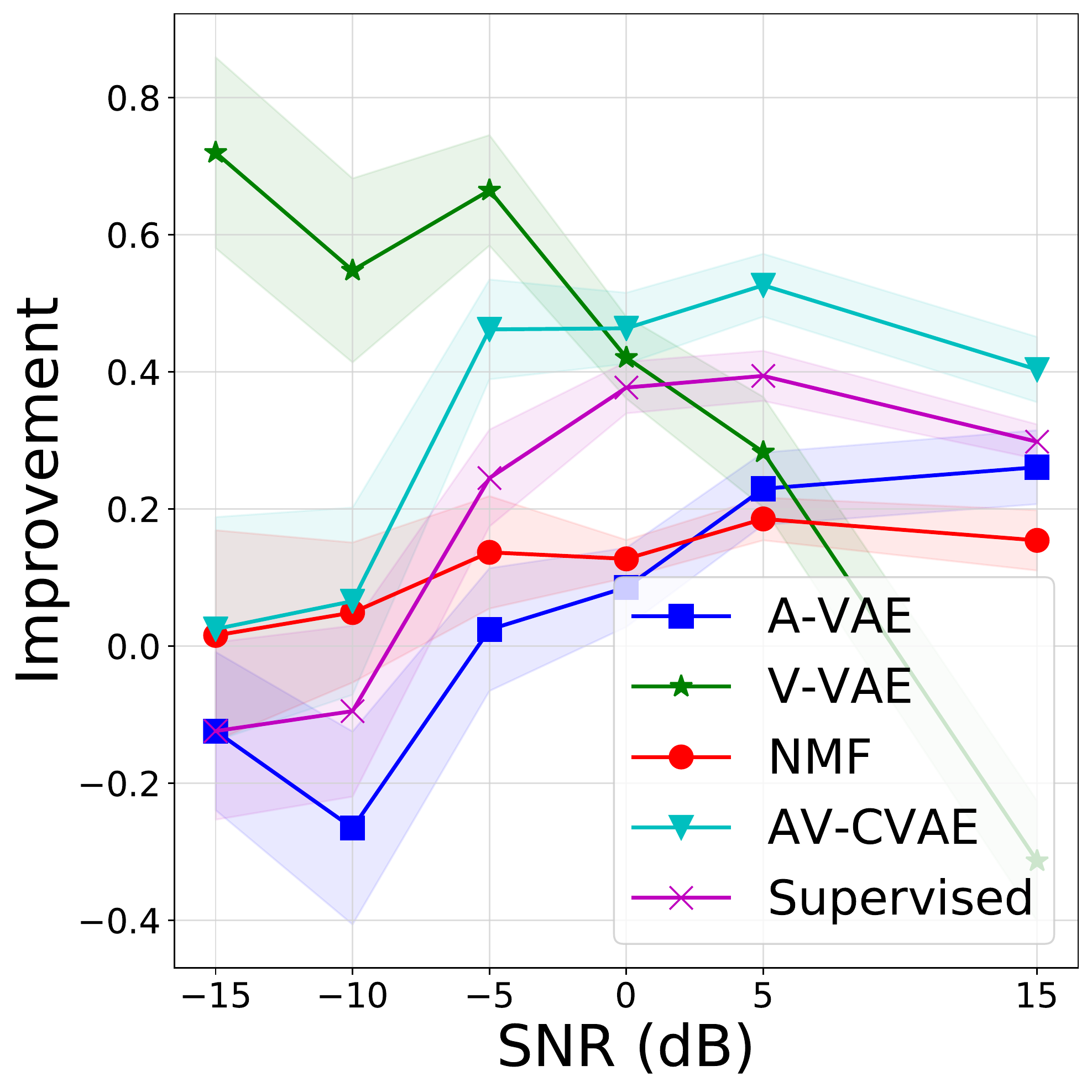} }}
	\subfloat[Cafe]{{\includegraphics[height=4cm]{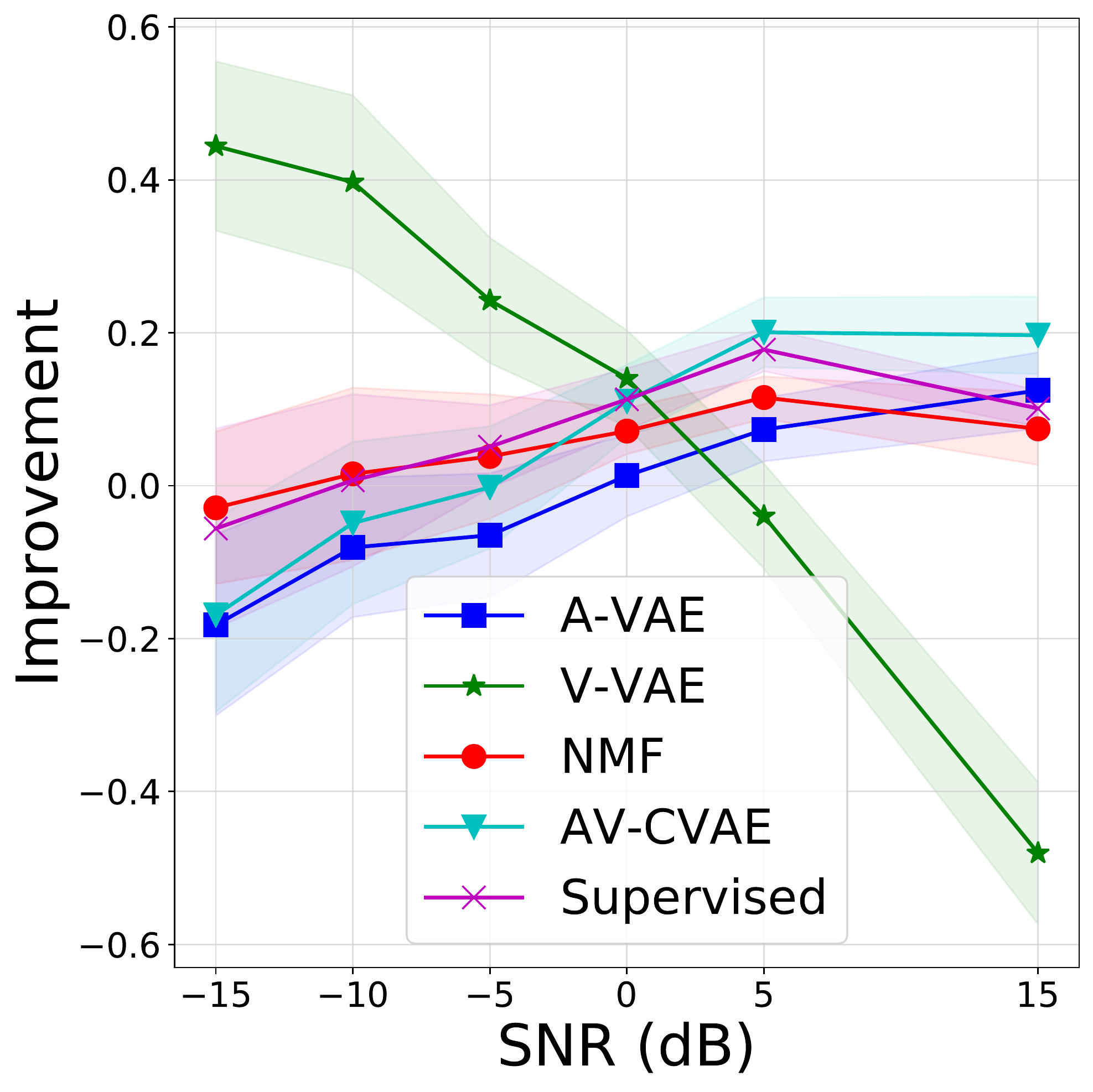} }}\\
	\subfloat[Car]{{\includegraphics[height=4cm]{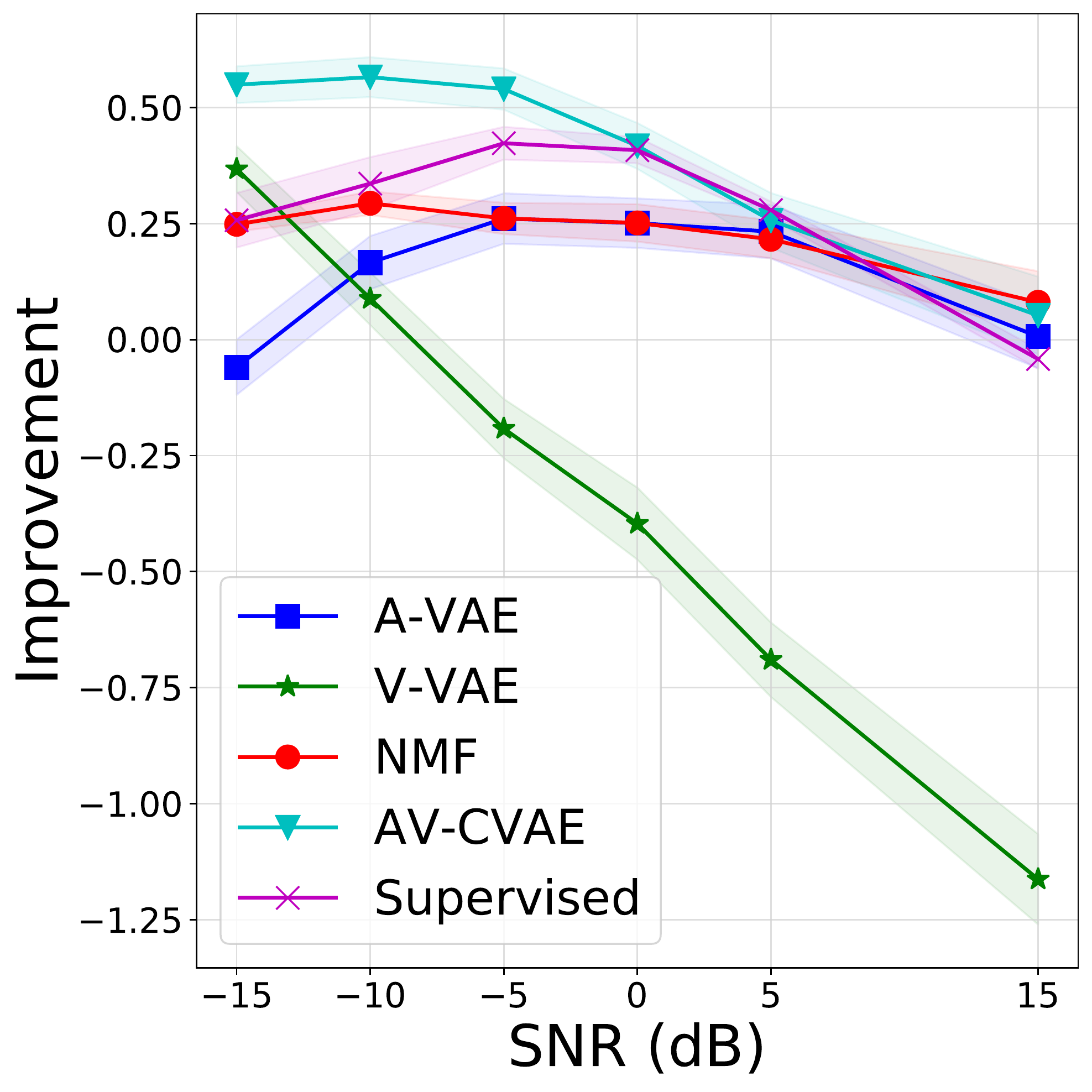} }}
	\subfloat[Babble]{{\includegraphics[height=4cm]{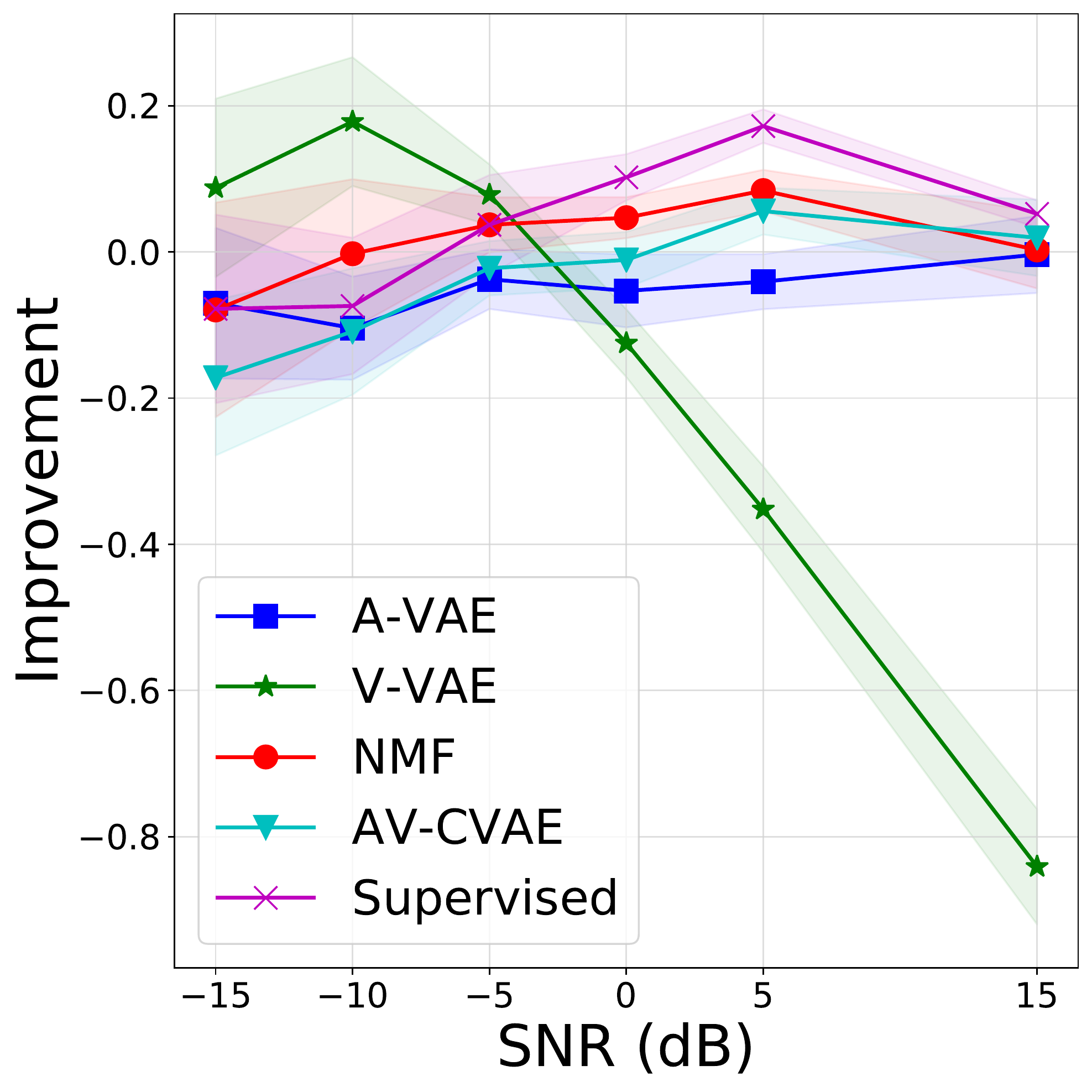} }}
	
	\caption{\label{fig:pesq_noise_types} PESQ results per noise types.}
\end{figure}

\begin{figure}[t!]
	\centering
	\subfloat[LR]{{\includegraphics[height=4cm]{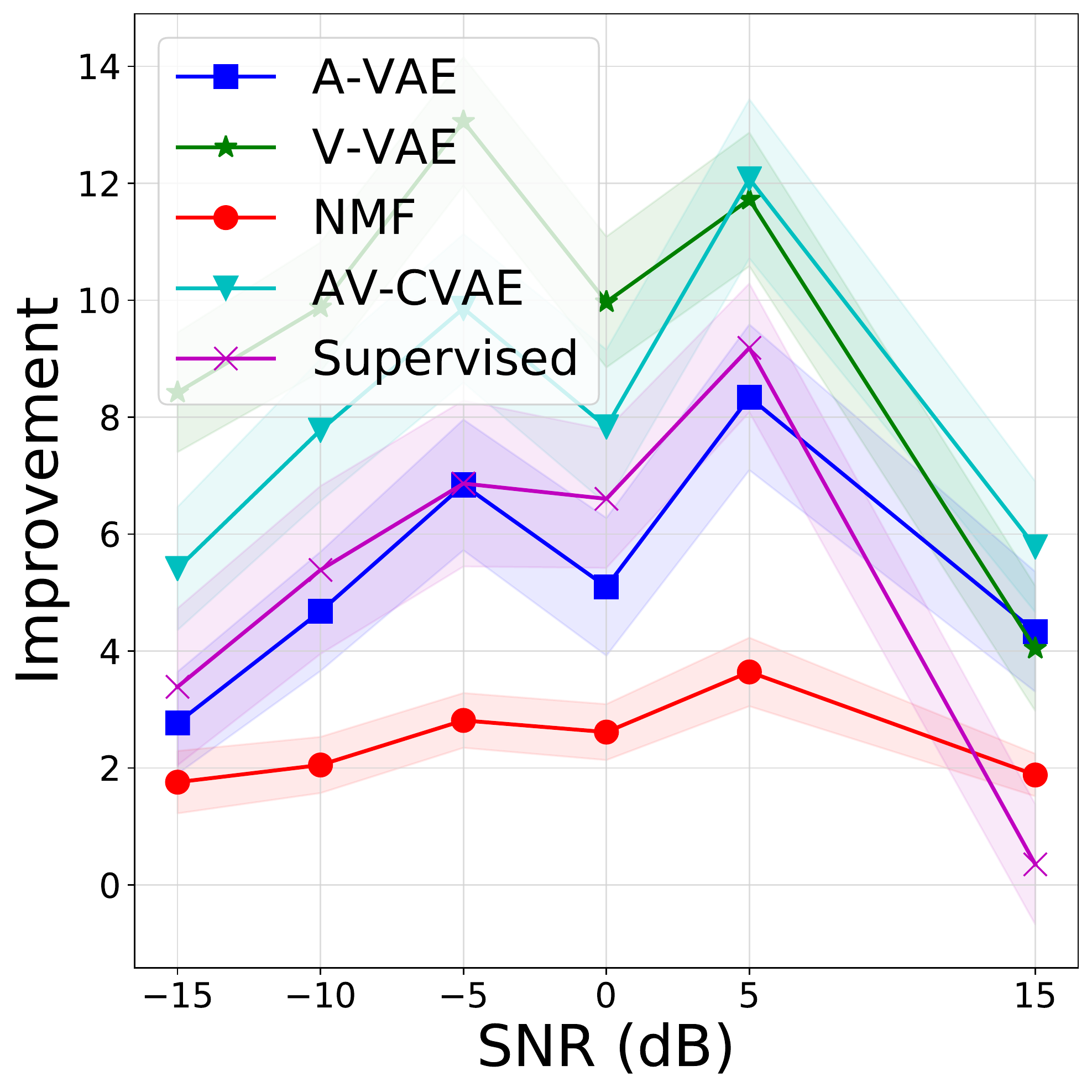} }}
	\subfloat[White]{{\includegraphics[height=4cm]{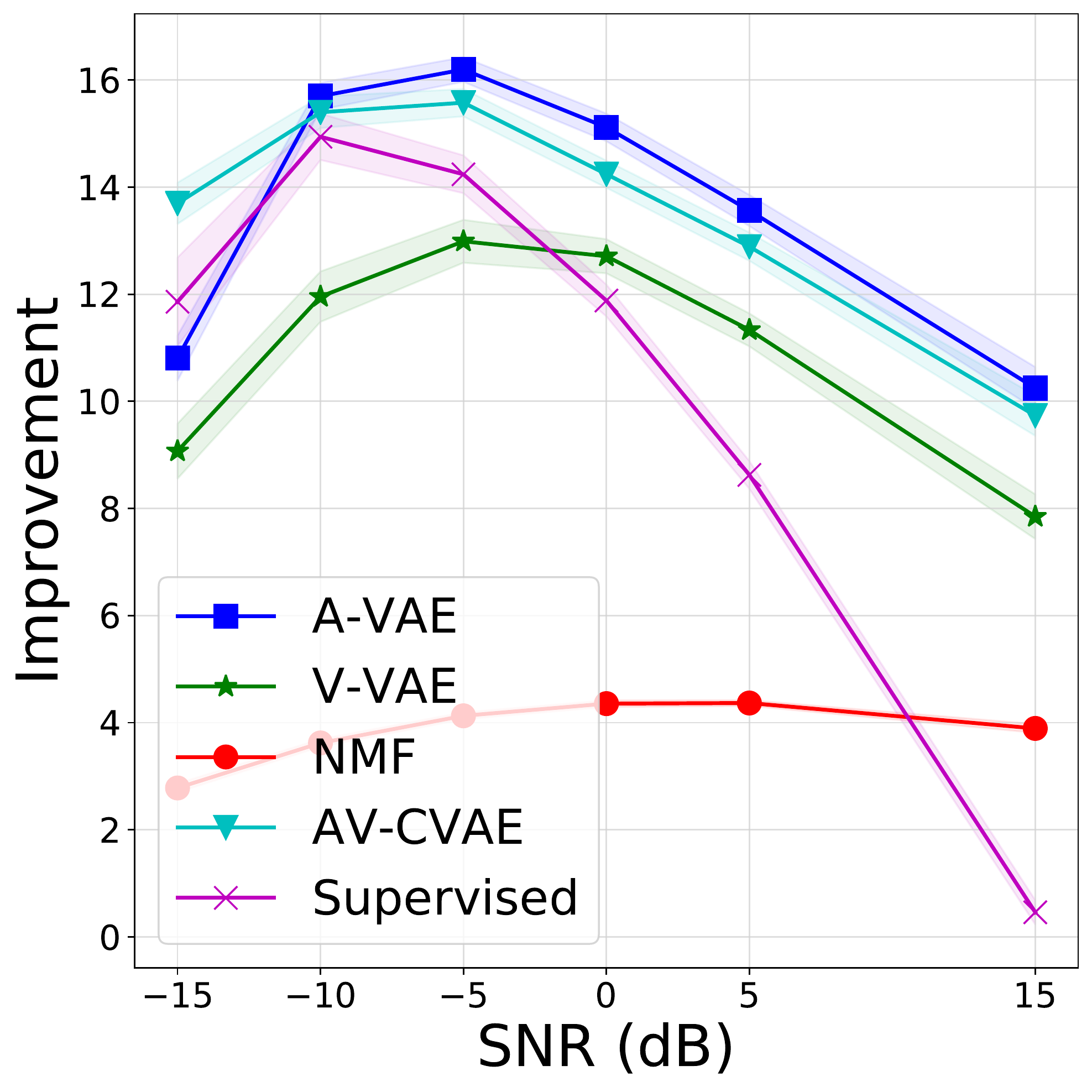} }}\\
	\subfloat[Street]{{\includegraphics[height=4cm]{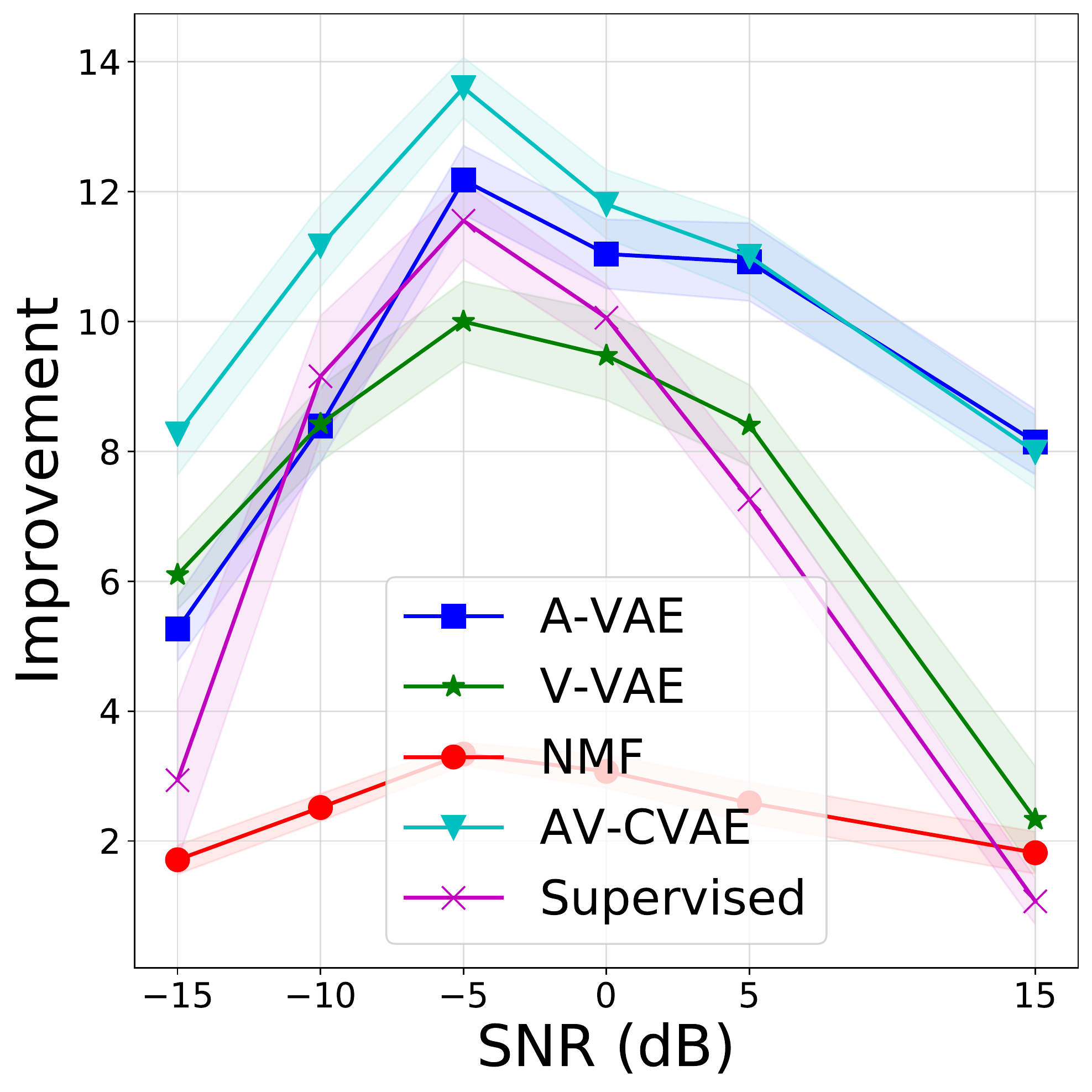} }}
	\subfloat[Cafe]{{\includegraphics[height=4cm]{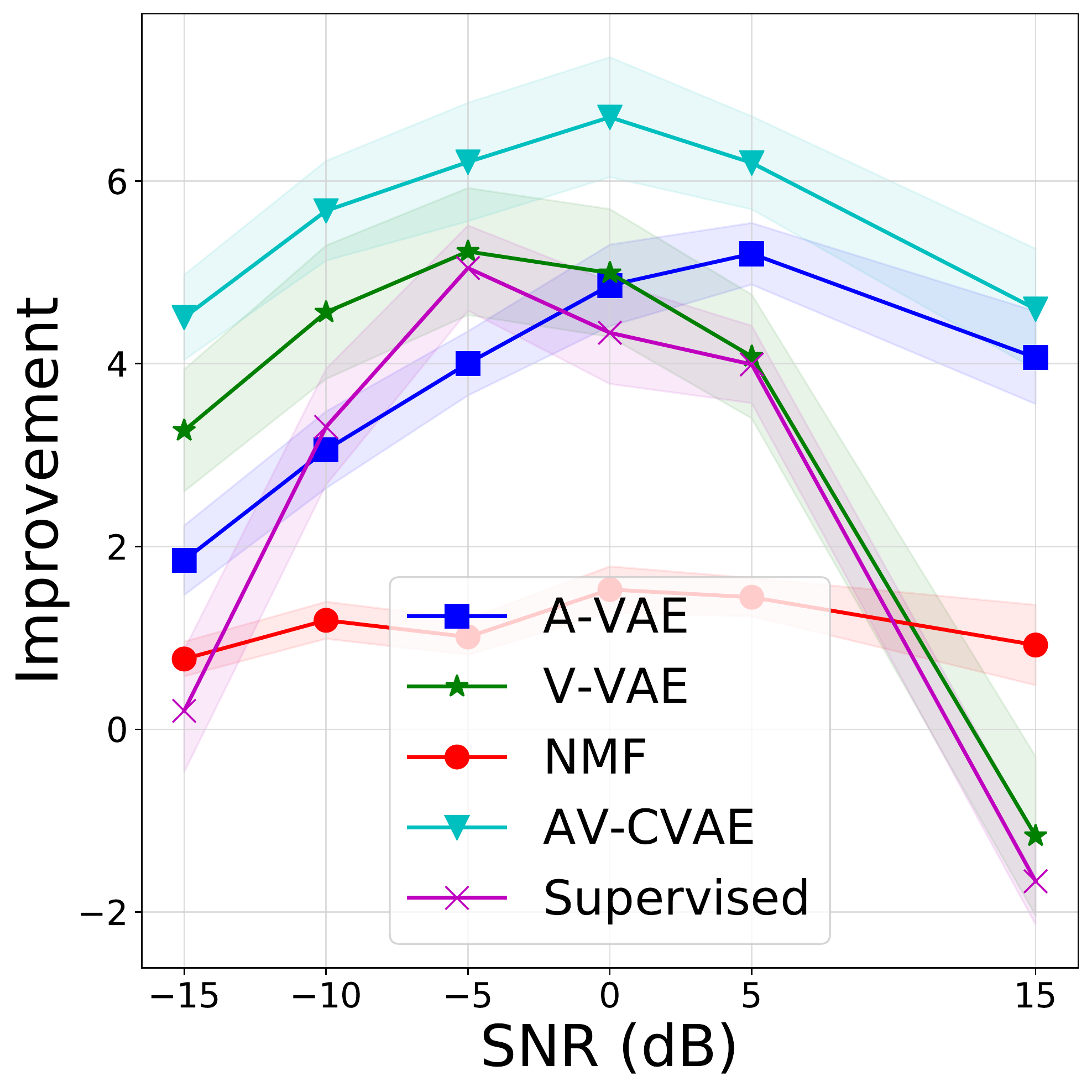} }}\\
	\subfloat[Car]{{\includegraphics[height=4cm]{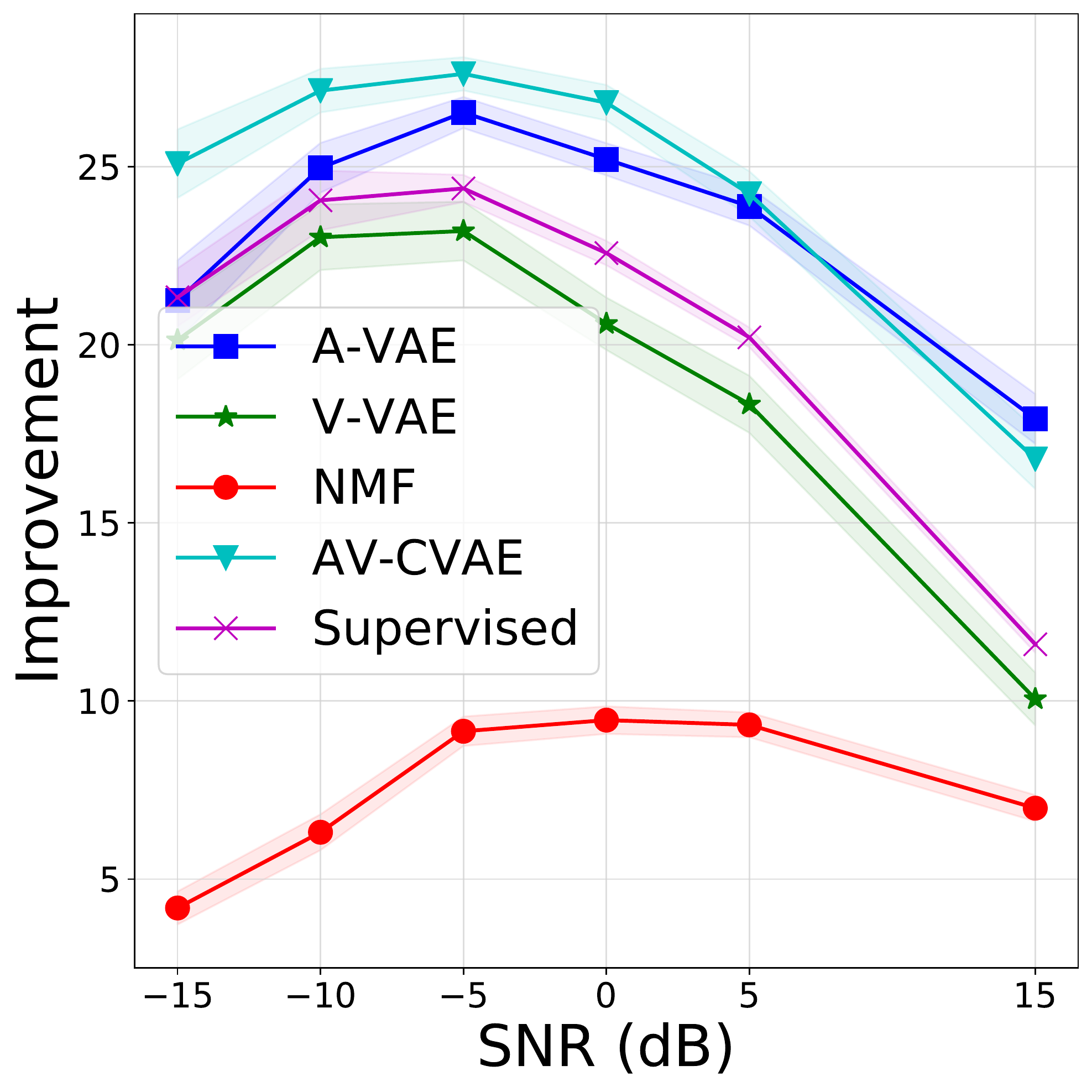} }}
	\subfloat[Babble]{{\includegraphics[height=4cm]{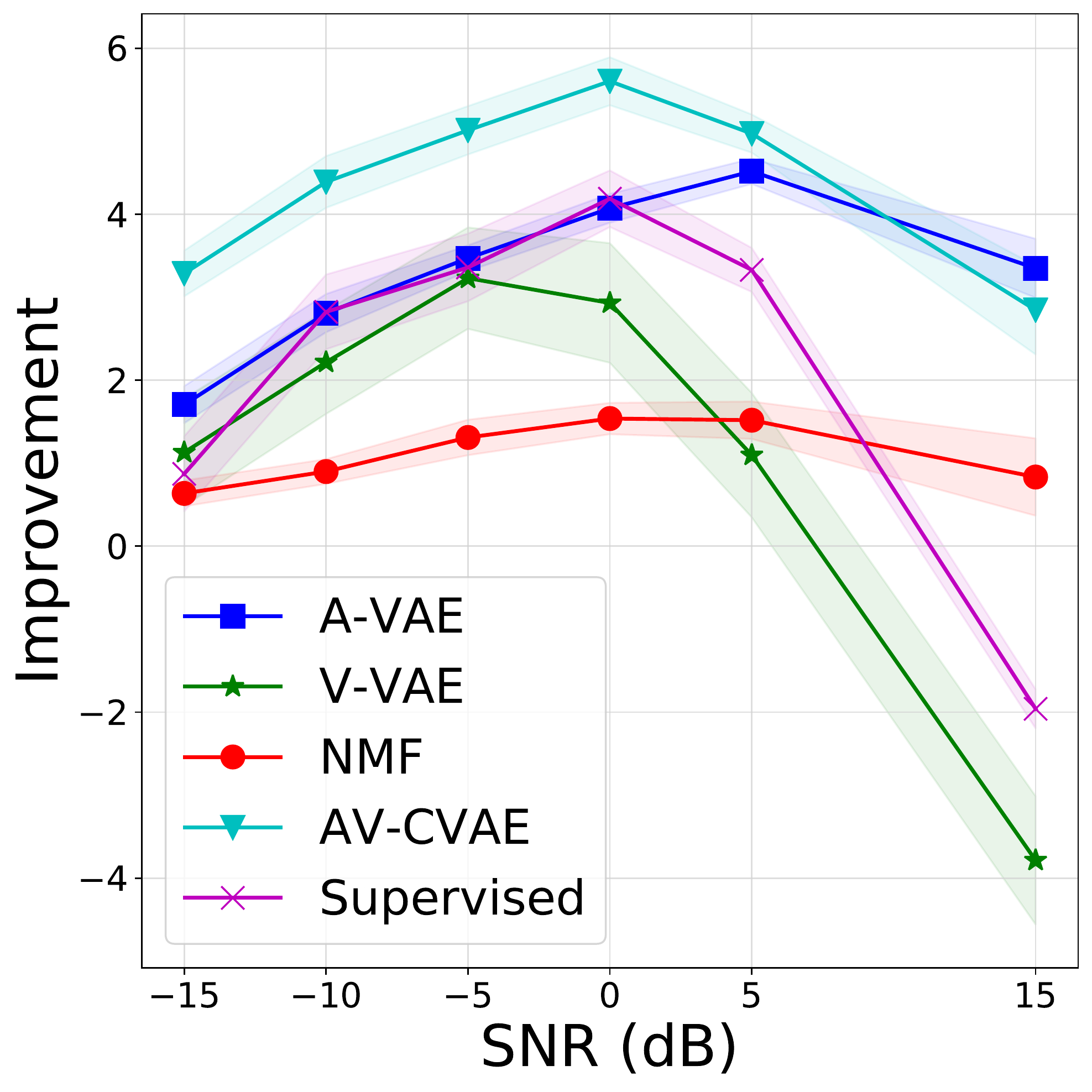} }}
	
	\caption{\label{fig:sdr_noise_types} SDR results per noise types.}
\end{figure}

We start by comparing the performance of A-VAE with the two V-VAE variants described in Section~\ref{sec:visual-vae} and Section~\ref{sec:vae-variants} and illustrated in Fig.~\ref{fig:v_vae}. The performance scores as a function of noise are shown in Fig.~\ref{fig:v_a}.
V-VAE performs better than A-VAE when the latter has to deal with high noise levels. One explanation for this could be the initialization of the latent variables in the Markov chain of the Metropolis-Hastings algorithm. In the case of V-VAE, the initialization is based on the visual features, whereas in A-VAE, it is based on the noisy mixture. Consequently, the former provides a better initialization than the latter, as it uses noise-free data (visual features). However, compared with A-VAE, the performance of V-VAE decreases as the noise level decreases. Intuitively, this is expected since the visual input of V-VAE does not depend on the noise associated with the audio signal. 
This intuition is confirmed by the curves plotted in
Fig.~\ref{fig:v_a}. \addnote[exp-v-vae]{1}{From this figure we can also see that the augmented V-VAE performs worse than the base V-VAE for high noise levels, especially in terms of PESQ. This can be explained by the fact that in V-VAE the posterior distribution of the latent variables is learned using only visual data. Therefore, by training from raw visual data, i.e., lip images, the network tries to learn as much as possible from raw visual data to reconstruct the audio data. Pre-trained visual features obtained from the ResNet, on the other hand, may not be as efficient as directly learning from raw visual data. This is because the ResNet has been trained to perform a different task.}

%

Next we assess the performance of AV-CVAE using two values for $ \alpha $ in \eqref{eq:cvae_elbotild}: $ \alpha=1 $ which corresponds to the original \ac{ELBO} in \eqref{eq:cvae_elbo}, 
$ \alpha=0.9 $ using the base V-VAE network and $ \alpha=0.85 $ using the augmented V-VAE. The score curves are plotted in 
Fig.~\ref{fig:cvae_bcvae} and one can see that the method performs better with  $ \alpha<1 $ than with $ \alpha=1 $. As explained in 
Section~\ref{sec:av_vae}, this improvement comes from the reduction of the gap  between the prior and the approximate posterior of the 
AV-CVAE model. In fact, the prior network is trained to generate latent vectors from visual features that are suitable for speech enhancement. 
In the following we thus use the AV-CVAE network with $ \alpha=0.9 $.

We also compared the performance of AV-CVAE with AV-VAE. We briefly remind that the former model uses visual information for training the prior, while the latter doesn't use visual information. Clearly, as it can be seen in Fig.~\ref{fig:vae_cvae}, AV-CVAE significantly outperforms AV-VAE, in particular with high noise levels.


In order to assess an overall performance of the proposed algorithms, we compared A-VAE, V-VAE, AV-CVAE, NMF \cite{SmarRS07_nmf}, and the state of the art method of 
 \cite{GabbSP18}. The scores plotted in Fig.~\ref{fig:all} show that AV-CVAE outperforms the A-VAE method by more than $ 2 $~dB, in terms of SDR, and by more than $ 0.3 $ in terms of the PESQ score. 
Moreover, AV-CVAE outperforms  \cite{GabbSP18} by more than $ 2 $dBs (on an average) in terms of SDR and by $ 0.1 $ in terms of PESQ. \addnote[exp-nmf1]{1}{Nevertheless, A-VAE and \cite{GabbSP18} outperforms AV-CVAE in terms of STOI for low noise levels. NMF, on the other hand, shows a worse performance than A-VAE and AV-CVAE. However, it outperforms A-VAE in terms of PESQ for high noise levels.}
\begin{figure}[t!]
	\centering
	\subfloat[LR]{{\includegraphics[height=4cm]{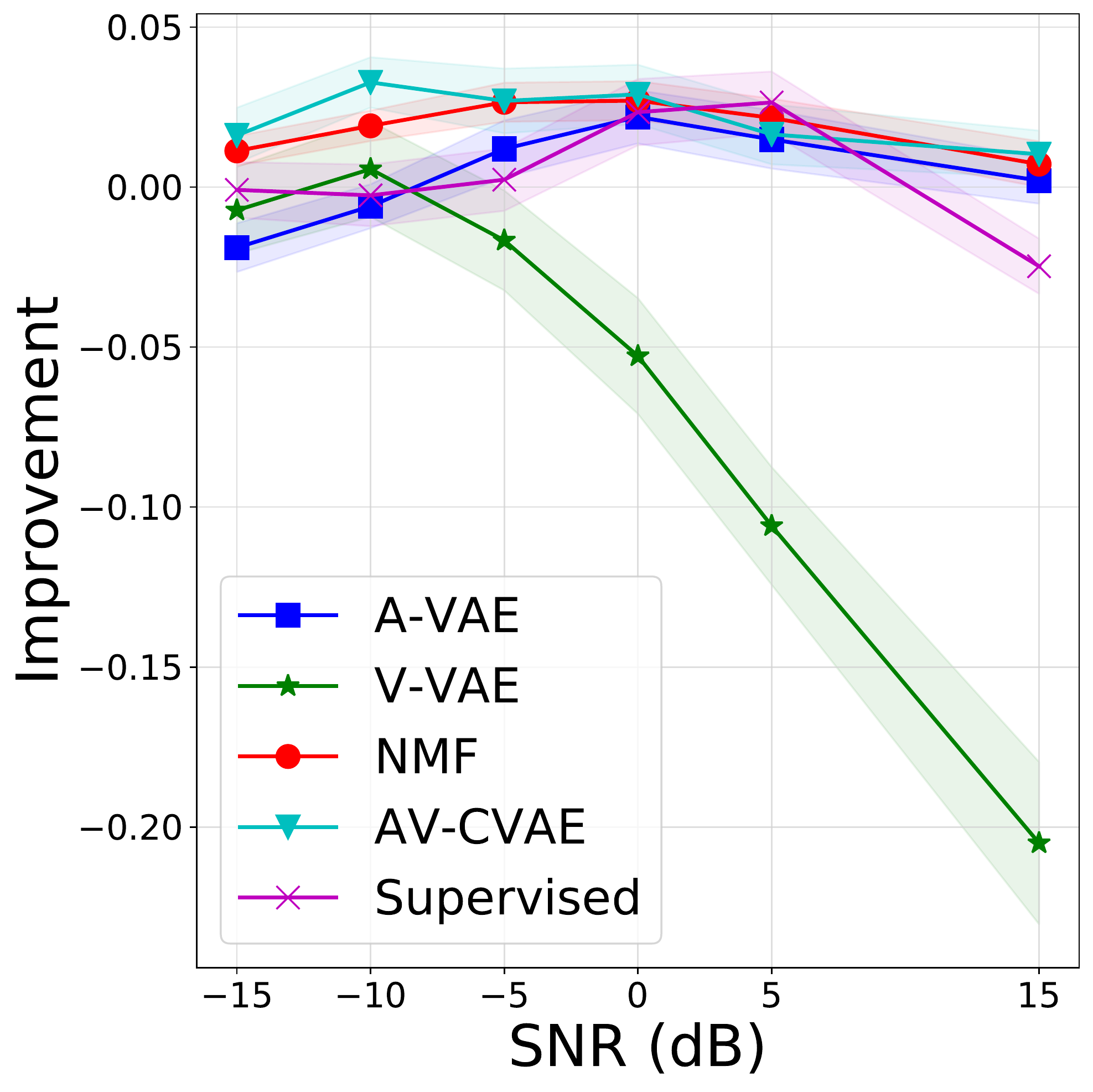} }}
	\subfloat[White]{{\includegraphics[height=4cm]{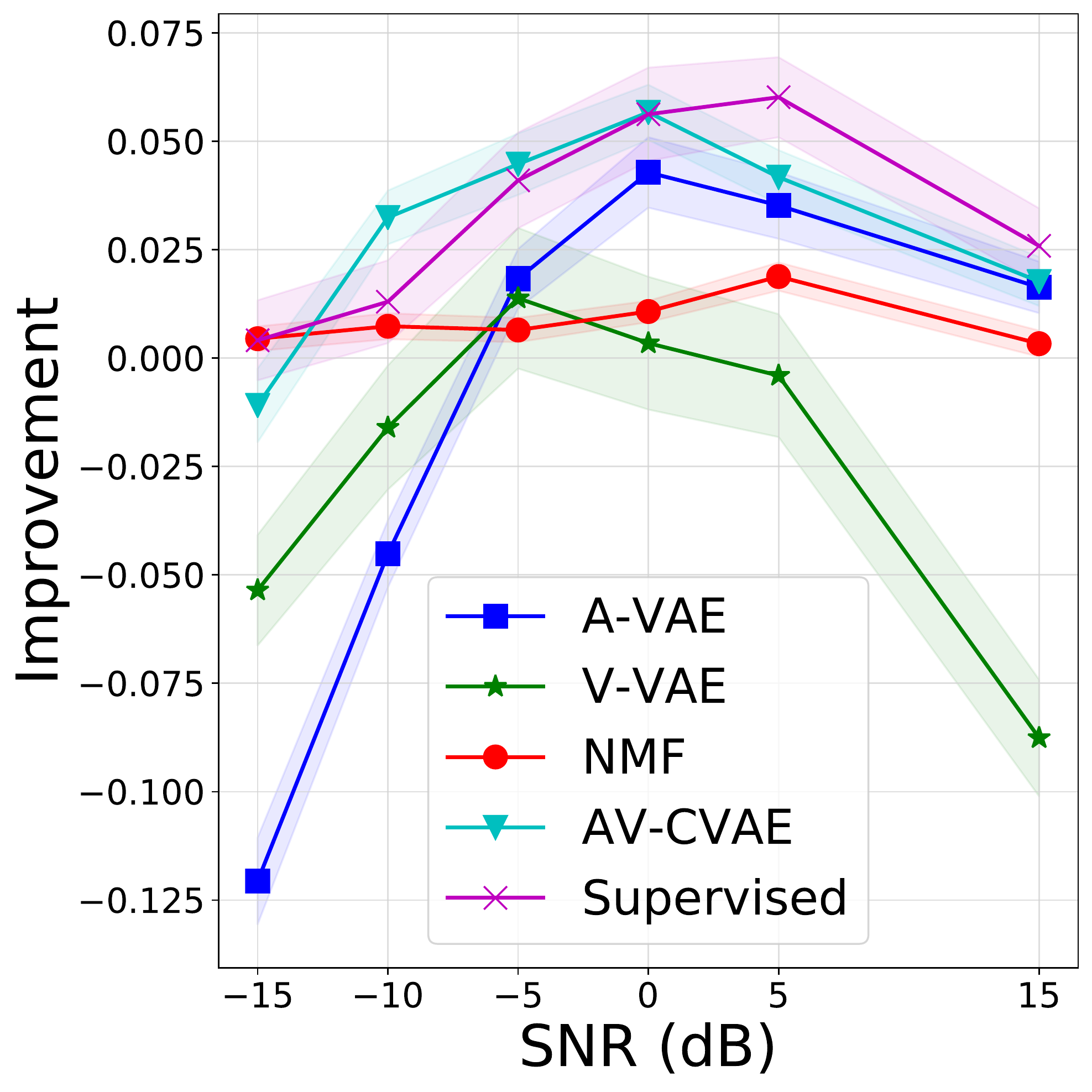} }}\\
	\subfloat[Street]{{\includegraphics[height=4cm]{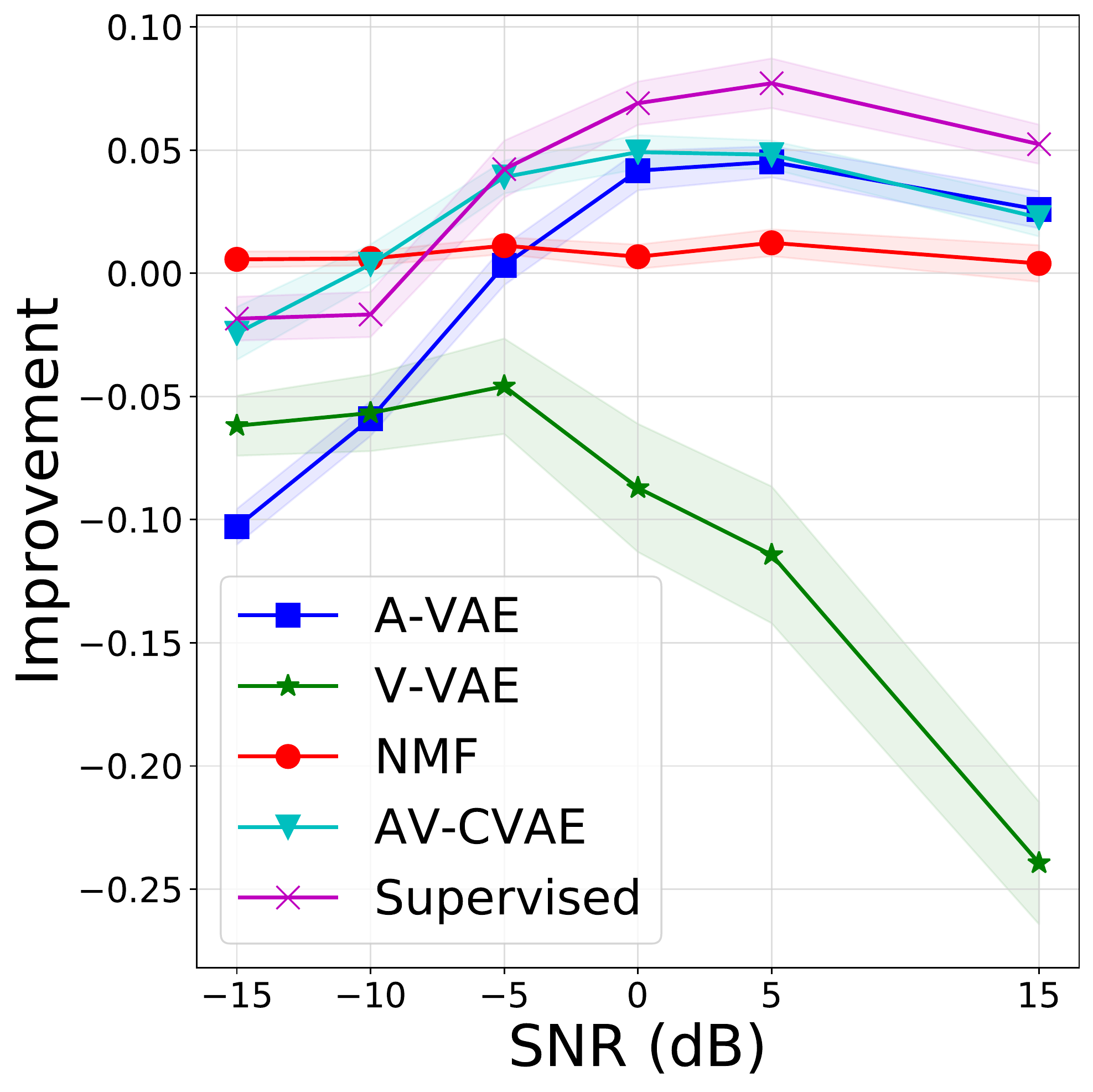} }}
	\subfloat[Cafe]{{\includegraphics[height=4cm]{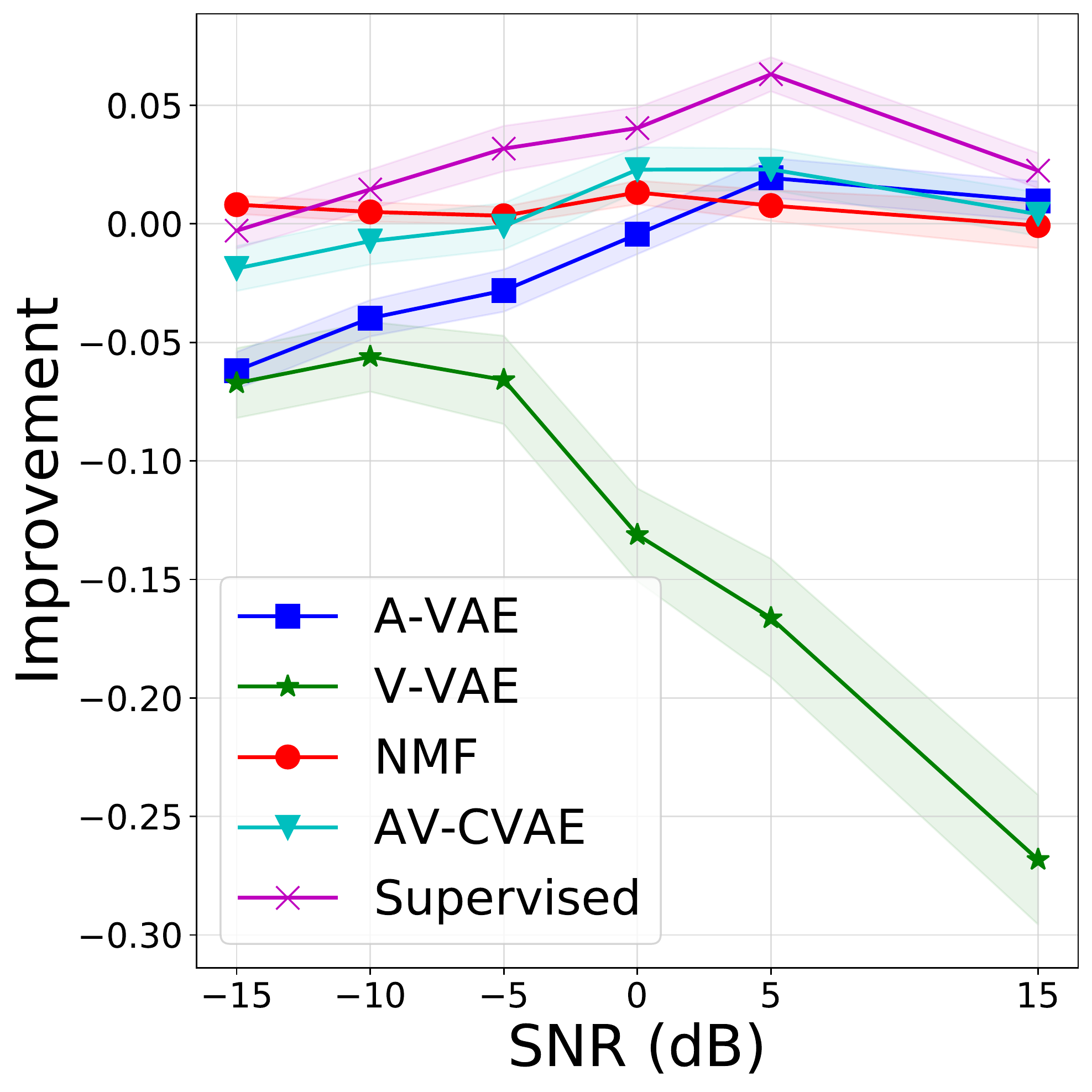} }}\\
	\subfloat[Car]{{\includegraphics[height=4cm]{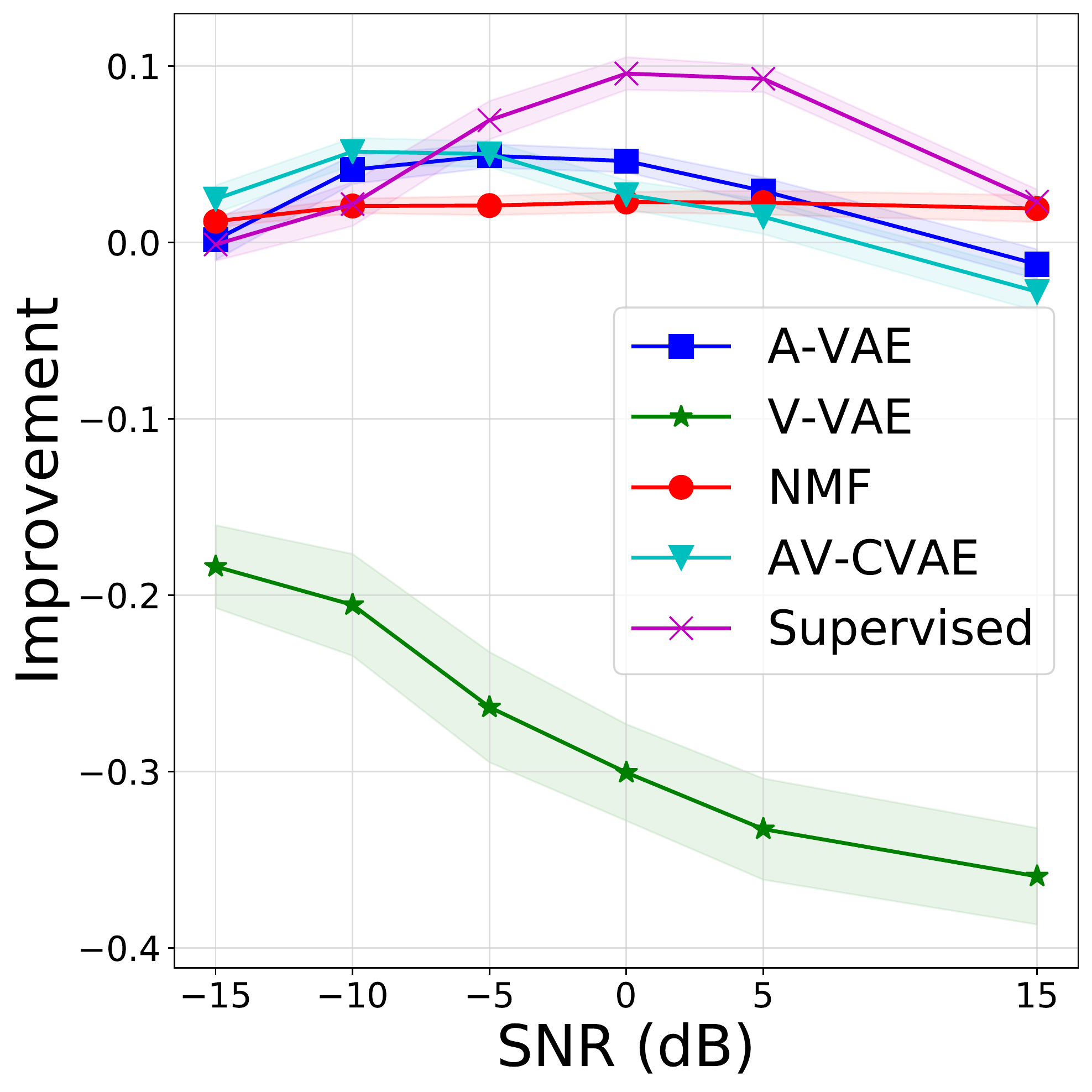} }}
	\subfloat[Babble]{{\includegraphics[height=4cm]{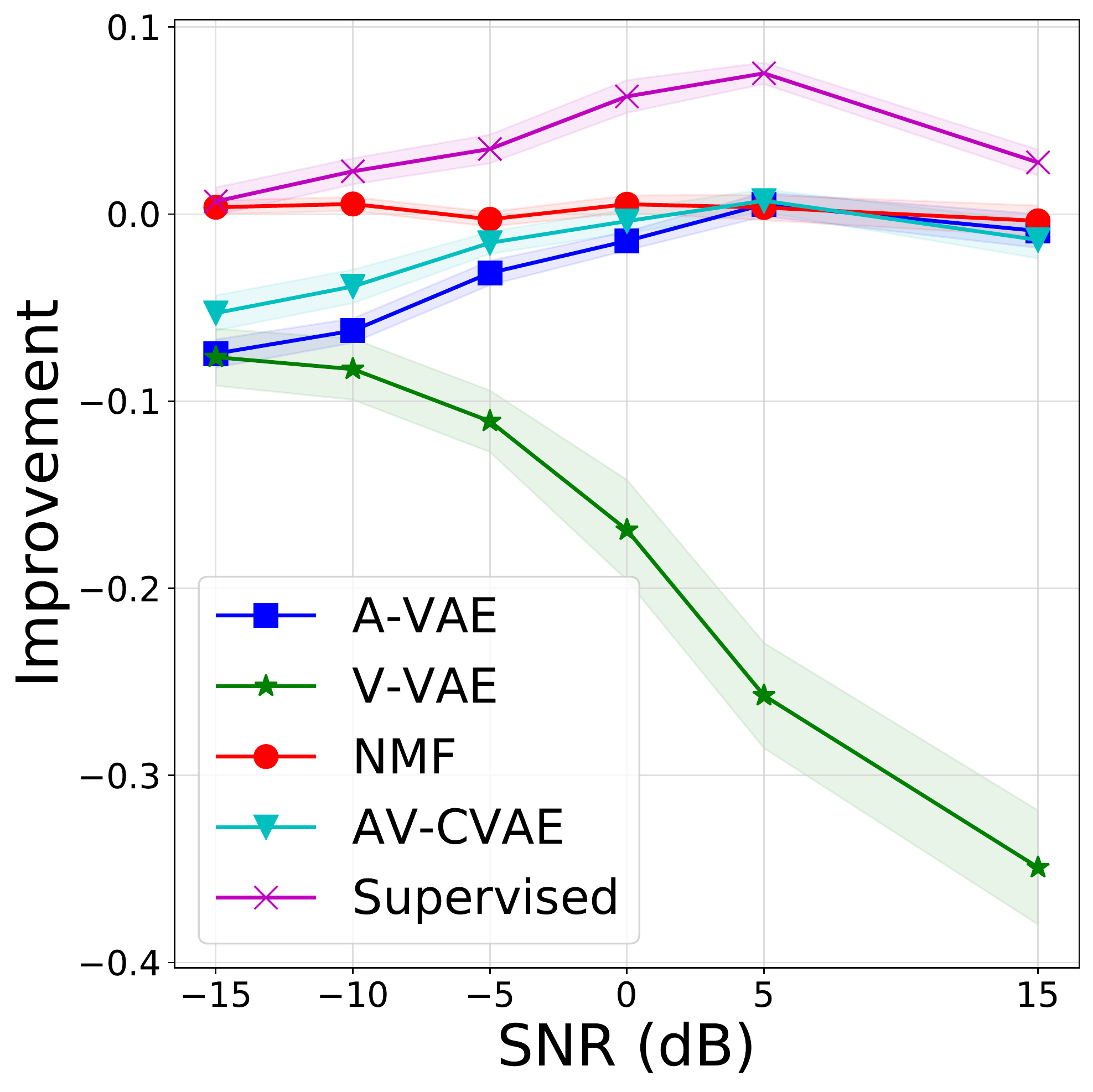} }}
	
	\caption{\label{fig:stoi_noise_types} STOI results per noise types.}
\end{figure}

\addnote[exp-nmf2]{1}{To see how the algorithms perform for each different noise type, we have reported the performance scores computed over all test samples, separately for each type of noise. Figures~\ref{fig:pesq_noise_types}, \ref{fig:sdr_noise_types}, and \ref{fig:stoi_noise_types} show the results. As can be seen, the performance ordering of the algorithms is different for each noise type.}

\begin{figure}[t!]
	\centering
	\subfloat[PESQ]{{\includegraphics[height=4.2cm]{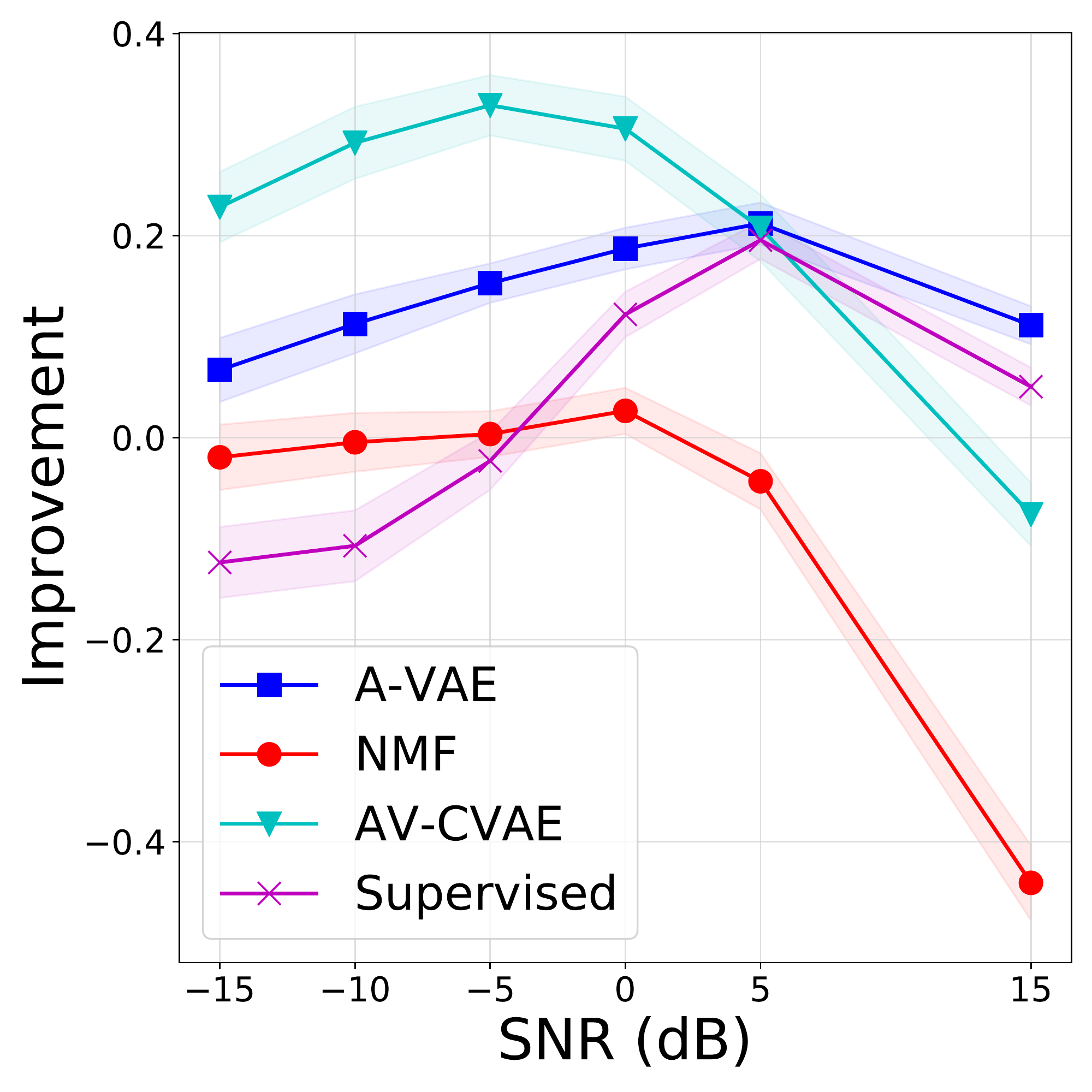} }}
	\subfloat[SDR]{{\includegraphics[height=4.2cm]{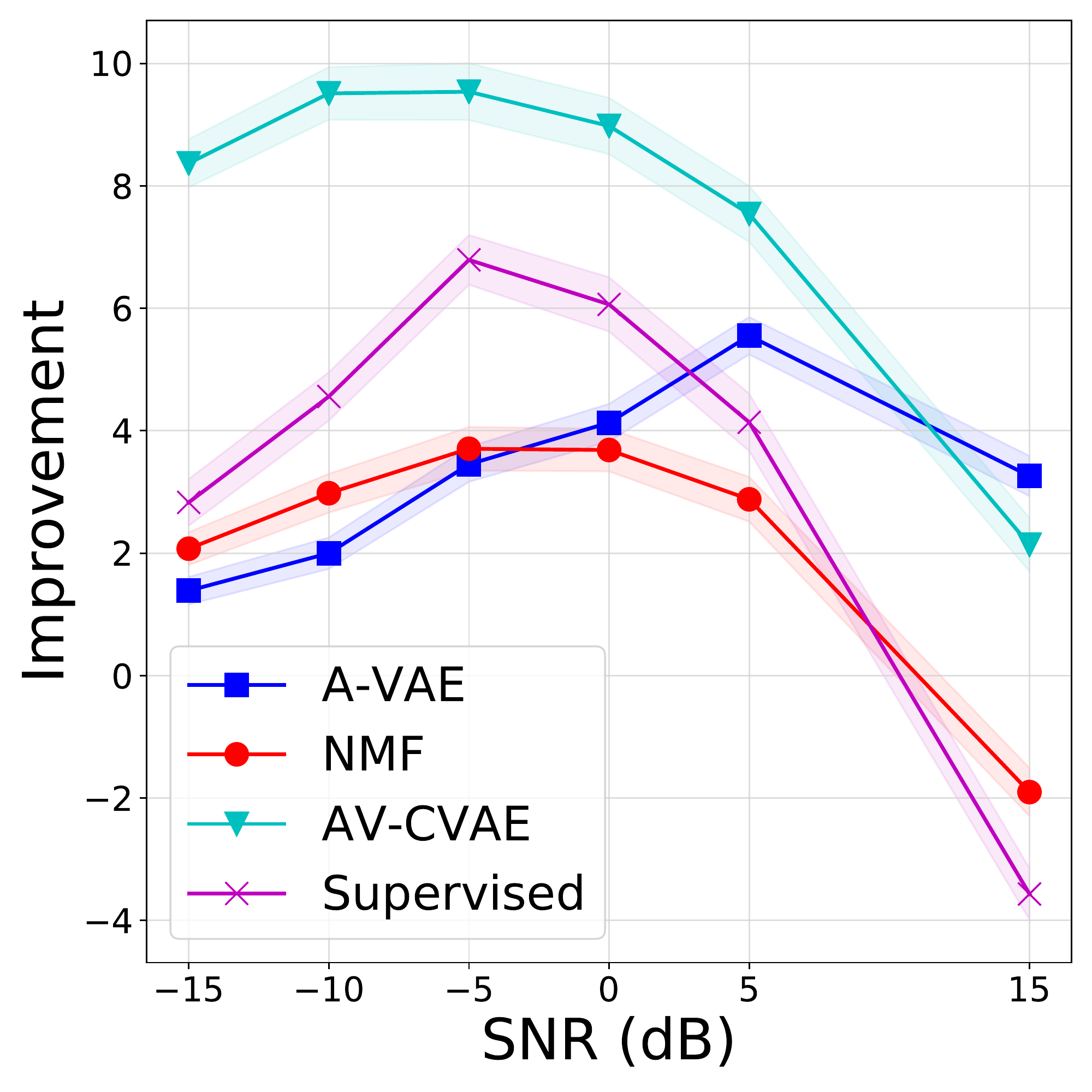} }}\\
	\subfloat[STOI]{{\includegraphics[height=4.2cm]{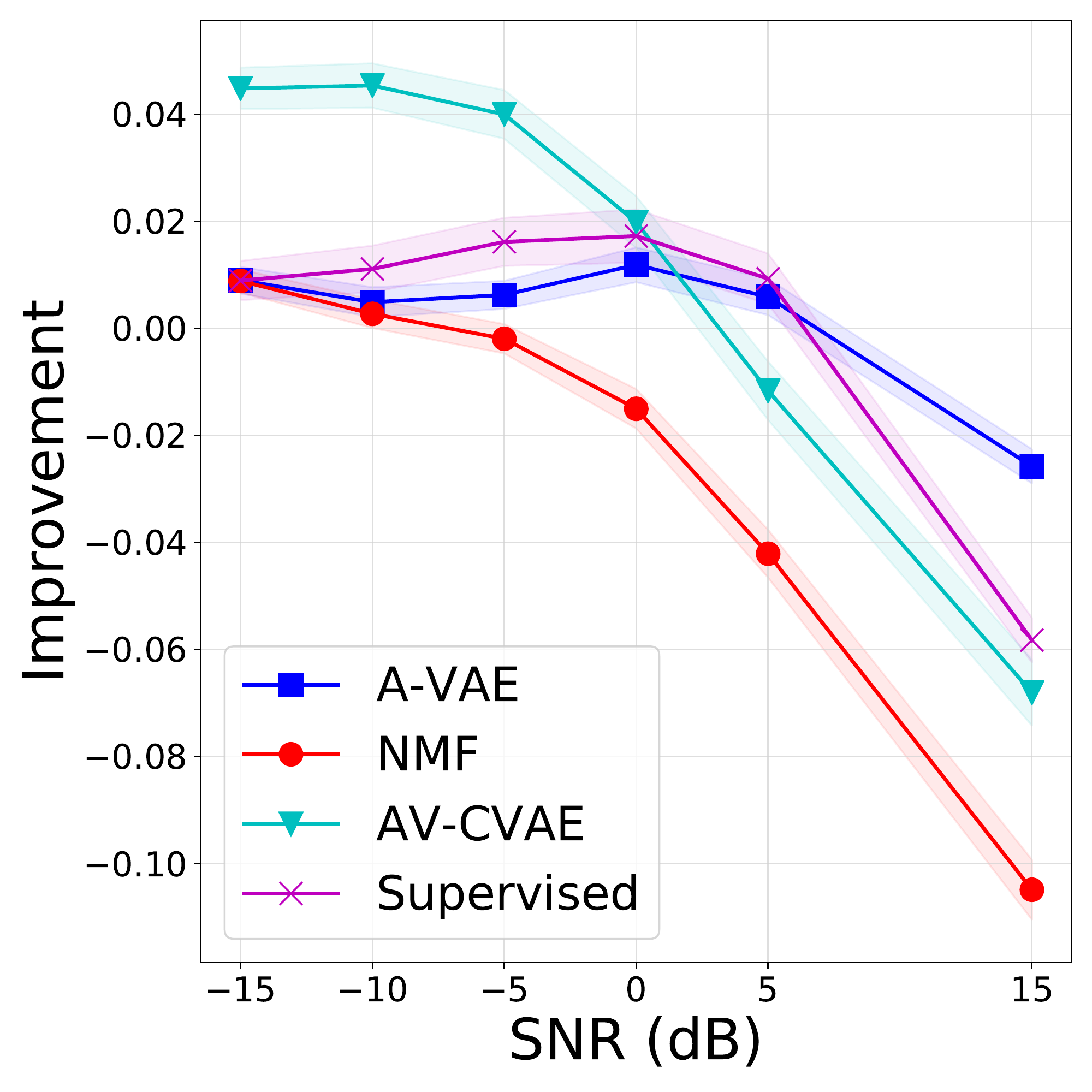} }}
	\caption{\label{fig:grid} Performance comparison of A-VAE, NMF,  AV-CVAE, and \cite{GabbSP18} on the GRID dataset. All the models have been trained on the NTCD-TIMIT dataset.}
\end{figure}

\addnote[exp-nmf3]{1}{Finally, we compared the performance of all the algorithms, trained using the NTCD-TIMIT dataset, on some test samples from the GRID dataset. The scores are shown in Fig.~\ref{fig:grid}, where we can see that AV-CVAE outperforms all the other methods in terms of PESQ (except for low noise levels, which is expected since visual information is less useful as the SNR is increased) and SDR with a significant margin.}

It should be emphasized that while the supervised method \cite{GabbSP18} needs to be trained with various noise types and noise levels, in order to have a good generalization performance, the proposed method is only trained on clean audio-speech and visual-speech  samples, independently of  noise types and noise levels. However, one advantage of \cite{GabbSP18} is its computational efficiency at testing. Also, notice that this method takes advantage of the dynamics of both the audio and visual data, through the presence of convolutional layers, which is not the case of our method that used fully-connected layers.

\section{Conclusions}
We proposed an audio-visual conditional VAE to model speech prior for speech enhancement. We described in detail several \ac{VAE} architecture variants and we provided details on how to estimate their parameters. We combined this audio-visual speech prior model with an audio mixture model and with a noise variance model based on \ac{NMF}. We derived an \ac{MCEM}
algorithm that infers both the time-varying loudness of the speech input and the noise variance parameters. Finally, a probabilistic Wiener filter performs speech reconstruction. 

Extensive experiments empirically validate the effectiveness of the proposed methodology to fuse audio and visual inputs for speech enhancement. In particular, the visual modality, i.e. video frames of moving lips, is shown to improve the performance, in particular when the audio modality is highly corrupted with noise.

Future works include the use of recurrent and convolutional layers in order to model temporal dependencies between audio and visual frames, and the investigation of computational efficient inference algorithms. It is also planned to extend the proposed \ac{AV-CVAE} framework to deal with more realistic visual information, e.g. in the presence of head motions and of temporary occlusions of the lips. Phase-aware speech generative models such as \cite{NugrahaICASSP2019} could also be considered for \ac{AV} speech enhancement.
\balance
\bibliographystyle{IEEEbib}

\end{document}